\documentclass{emulateapj}
\newcommand{\etal}{et al.\ }
\newcommand{\etalb}{et al.}
\newcommand{\be}{\begin{equation}}
\newcommand{\ba}{\begin{eqnarray}}
\newcommand{\ee}{\end{equation}}
\newcommand{\ea}{\end{eqnarray}}

\begin{document}
\title{Orbital structure of merger remnants: Trends with gas fraction in 1:1 mergers}

\author{Loren Hoffman\altaffilmark{1}, Thomas
J. Cox\altaffilmark{2,3}, Suvendra Dutta\altaffilmark{2,4},
Lars Hernquist\altaffilmark{2}}

\email{l-hoffman@northwestern.edu}

\altaffiltext{1}{Department of Physics and Astronomy, Northwestern
University, Dearborn Observatory, 2131 Tech Drive, Evanston, IL 60208}

\altaffiltext{2}{Department of Astronomy, Harvard University, 60 Garden Street, Cambridge, MA 02138}

\altaffiltext{3}{Carnegie Observatories, 813 Santa Barbara Street, Pasadena, CA 91101}

\altaffiltext{4}{Faculty of Arts and Sciences, Harvard University, University Hall, Cambridge, MA 02138}

\begin{abstract}
Since the violent relaxation in hierarchical merging is incomplete, elliptical galaxies retain a wealth of information about
their formation pathways in their present-day orbital structure.  Recent advances in integral field spectroscopy, multi-slit 
infrared spectroscopy, and triaxial dynamical modeling techniques have greatly improved our ability to harvest this 
information.  A variety of observational and theoretical evidence indicates that gas-rich major mergers play an important 
role in the formation of elliptical galaxies.  We simulate 1:1 disk mergers at seven different initial gas fractions 
($f_{gas}$) ranging from 0 to 40\%, using a version of the TreeSPH code Gadget-2 that includes radiative heating and cooling, 
star formation, and feedback from supernovae and active galactic nuclei.  We classify the stellar orbits in each remnant and 
construct radial profiles of the orbital content, intrinsic shape, and orientation.  The dissipationless remnants are 
typically prolate-triaxial, dominated by box orbits within $r_{c} \sim 1.5R_{e}$, and by tube orbits in their outer parts.  
As $f_{gas}$ increases, the box orbits within $r_{c}$ are increasingly replaced by a population of short axis tubes 
($z-$tubes) with near zero net rotation, and the remnants become progressively more oblate and round.  The long axis tube 
($x-$tube) orbits are highly streaming and relatively insensitive to $f_{gas}$, implying that their angular momentum is 
retained from the dynamically cold initial conditions.  Outside $r_{c}$, the orbital structure is essentially unchanged by 
the gas.  For $f_{gas} \gtrsim 15$\%, gas that retains its angular momentum during the merger re-forms a disk, that 
appears in the remnants as a highly streaming $z-$tube population superimposed on the hot $z-$tube distribution formed by the 
old stars.  In the 15-20\% gas remnants, this population appears as a kinematically distinct core (KDC) within a system that 
is slowly rotating or dominated by minor-axis rotation.  These remnants show an interesting resemblance, in both their 
velocity maps and intrinsic orbital structure, to the KDC galaxy NGC4365 \citep{ngc4365}.  At 30-40\% gas, the remnants 
are rapidly rotating, with sharp embedded disks on $\sim 1R_{e}$ scales.  We predict a characteristic, physically intuitive 
orbital structure for 1:1 disk merger remnants, with a distinct transition between 1 and 3$R_{e}$ that will be readily 
observable with combined data from the 2D kinematics surveys SAURON and SMEAGOL.  Our results illustrate the power of 
direct comparisons between $N-$body simulations and dynamical models of observed systems to constrain theories of galaxy 
formation. 
\end{abstract}


\section{Introduction}

\subsection{Motivation}

In the standard $\Lambda$CDM concordance cosmology \citep{os95,dod96,wmap3}, structure 
in the Universe grows hierarchically, through a progression of smaller bodies accreting
material and merging to form larger systems (e.g. \citealt{wr78}).  Cosmological $N$-body simulations 
starting from a Gaussian random field of linear density fluctuations at high redshift 
(e.g. \citealt{millsim}), together with analytic models of gravitational collapse 
(e.g. \citealt{bertsch}), have given us a fairly clear picture of how dark matter (DM) 
assembles in the Universe.  Extended Press-Schechter theory \citep{ps,bond,lc93} provides 
an analytic formalism for computing halo merger rates and assembly histories, which matches 
$N$-body simulations remarkably well \citep{lc94,shy09}, and has been used in a variety of 
semianalytic models of structure formation 
(e.g. \citealt{colesemi,ms96,kaufsemi,somsemi,crotsemi,som2semi,s09b}).  
Simulated DM halos also appear to share a universal internal morphology, with density and 
velocity anisotropy profiles similar to the generic outcome of violent relaxation following 
dissipationless collapse or strong tidal shocking 
\citep{dc91,nfw,bull01a,bull01b,nav08,ms79,va82,mg84,mg90,sh92,huss99,mac06,bell08}. 

There is no such simple theory of hierarchical {\it galaxy} formation, because the luminous
components of galaxies are formed through complex baryonic physics.  For instance radiative 
heating and cooling, star formation (SF) and gas expulsion through stellar winds, energy and
momentum feedback from supernovae and active galactic nuclei (AGN), ram pressure stripping, 
and resonant effects in dynamically cold systems, are all important ingredients in determining 
the baryonic structure of galaxies 
(e.g. \citealt{crotsemi,sprsf,gadfeedbak,best07,tjfeed,cevklyp,reproc,dong09}).  Cosmological models 
of galaxy formation must therefore be calibrated with high-resolution simulations on galactic 
scales that study these effects in isolation, coupled with direct constraints from observations.  

Violent relaxation in mergers tends to drive galaxies toward a universal, fully mixed 
structure \citep{all65,l67,sw98,th09}, while gas accretion and stellar outflows produce 
new dynamically cold components 
\citep{rw90,block02,shy08,dekel09,bourelme,reproc,gasz15}.  
From the hot stellar distributions of elliptical galaxies, we deduce 
that they are the most merger-dominated systems.  Ellipticals and bulges contain the 
majority of the stellar mass in the local Universe (e.g. \citealt{g09bulges}), and often 
serve as a testing ground for theory since they are the most evolved under the complex 
combination of processes driving galaxy formation.  Since the violent relaxation in mergers 
is incomplete, elliptical galaxies retain a wealth of information about their formation 
histories in their present-day distribution functions (e.g. \citealt{egg62,l67,w80mix,v07relax}).  

Recent advances in integral field spectroscopy (IFS; \citealt{tiger,sauron1,virus,saur4re}), 
multi-slit infrared spectroscopy \citep{norr08,proc09}, and triaxial dynamical modeling techniques 
\citep{schwarzmod,ngc4365,nmagic,nmagic3} have greatly improved our ability to harvest this information.  
The SAURON project \citep{sauron1,sauron3} will produce high resolution, 2D kinematic maps within $\sim 1$ 
effective radius ($R_{e}$) for a representative sample of $\sim$100 nearby elliptical galaxies 
and spiral bulges, using a panoramic integral field spectrograph mounted on the William Herschel 
Telescope.  The data on 48 early-type galaxies released to date has revealed an unexpectedly 
rich variety of kinematic structures, which poses a new challenge for galaxy formation 
simulations \citep{jess07,bur08}.  Dynamical modeling studies have shown that, 
in practice, 2D maps of the first four moments ($h_{1}-h_{4}$) of the line-of-sight velocity 
distribution (LOSVD) provided by SAURON are typically sufficient to uniquely reconstruct the 
3D stellar orbital distribution \citep{schwarzmod,ngc4365}.  Complex features present in many systems, 
such as embedded disks and kinematically distinct cores (KDCs), can provide especially strong 
constraints on the intrinsic structure \citep{vv08}.  

A good example of the power of these new techniques is provided by the case of NGC4365.  
This massive old elliptical is known for its minor axis rotation \citep{w88,b94} and KDC \citep{ngc4365kdc}, 
and is therefore a natural candidate for dynamical modeling. 
\citet{stat4365} modeled this galaxy using a velocity field fitting method \citep{statshapes1,statshapes3} that 
made use of the surface brightness and full 2D velocity map from SAURON, but not the higher 
moments of the LOSVDs. They found that the system was nearly maximally triaxial, ruling out axisymmetry 
at $>$95\% confidence. 

A few years later, \citet{ngc4365} modeled the same galaxy using an advanced new 
triaxial Schwarzschild modeling \citep{schw79} code that can incorporate all of the LOSVD 
moments up to $h_{4}$. They reached a qualitatively different conclusion - that the system was 
nearly oblate axisymmetric.  The predominance of the minor-axis rotation in the outer 
parts of the map owed to a high degree of cancellation of the orbits rotating in a 
prograde and retrograde sense about the short axis, and streaming of the smaller population 
of orbits rotating about the intrinsic long axis.  They found no major transition in the 
orbital structure at the boundary of the kinematically ``decoupled'' core, making it unlikely 
that it formed in a separate infall event.  The orbital structure of the 15-20\% gas merger 
remnants in this paper bears a tantalizing resemblance to that of NGC4365 (compare e.g. our 
Figures~\ref{fig:kdcs} -~\ref{fig:kdcspec} and~\ref{fig:rems15} -~\ref{fig:rems20} with 
Figures 7, 11 and 12 of \citealt{ngc4365}).

However a limitation of the SAURON spectrograph is its small field of view, corresponding to $\sim 1R_{e}$ 
on a typical elliptical target.  The outer parts of galaxies are less relaxed than their inner parts, 
and bar-like modes in mergers efficiently transport angular momentum outward 
\citep{op73,fall79,hernquist93,bh96,hop08b,hoff09c}.
A gas-rich merger between two spiral galaxies with halos might be expected to produce a remnant with
three distinct components in its distribution function: (i) an inner part formed through dissipation,
(ii) a middle part reflecting the dynamically cold distribution of the disk stars, and (iii) an outer
part arising from the pre-existing stellar halo populations.  Observations with about four times the spatial
coverage of SAURON would be able to detect these dynamical subcomponents, and probe the parts of galaxies 
retaining the most memory of their progenitors' angular momentum and internal structure.  Hints of increased
complexity in the angular momentum profiles at large radii have indeed been observed in a few elliptical
galaxies \citep{cocc09,proc09}.

A number of projects designed to extend SAURON-style dynamical modeling out to larger radii are
currently underway.  The SMEAGOL survey \citep{proc09,fost09} will obtain smoothed 2D maps of $h_{1}-h_{4}$
out to $\sim 3R_{e}$ for a representative sample of 25 nearby ellipticals, using the new 
stellar kinematics with multiple slits (SKiMS) technique \citep{norr08,proc08,proc09} with 
the DEIMOS spectrograph on the 10-meter Keck-II telescope.  The data analysis will include triaxial 
dynamical modeling with the advanced particle-based method NMAGIC 
\citep{madetomeas,nmagic,nmagic2,nmagic3}.  The wide-field IFS VIRUS-P has been used to obtain 
2D stellar kinematics out to $3-4R_{e}$ for several giant ellipticals \citep{virus,blanc09,virus2}, 
and multiple pointings of the SAURON spectrograph have been used to measure the stellar LOSVDs
out to $\sim 4R_{e}$ in NGC3379 and NGC821 \citep{saur4re}.  Surveys using globular clusters (GCs)
and planetary nebulae (PNe) as discrete tracers of the distribution function 
\citep{sages,doug02,doug07,nap09,nantais09,schub09} can probe much larger radii (out to $\sim 10R_{e}$) and 
place further constraints on dynamical models \citep{discrete}.

Combined with simulations aimed at establishing the characteristic orbital structure arising
from various formation pathways, these observational programs will provide unprecedented
insight into the physics of galaxy formation.

\subsection{Orbits in galactic potentials}

To deduce the formation histories of galaxies from their orbital structure, we must identify macroscopic 
groups of orbits that are confined to a well-defined neighborhood of phase space because they followed
similar evolutionary pathways.  The isolating integrals of motion (or quasi-isolating integrals in the case of
perturbed potentials; \citealt{c63b,gs81,siopkan00,kansiop03}) parameterize the phase space region in which a 
star remains localized, and encode whatever information about its initial conditions (ICs) is preserved once the 
system is fully phase-mixed (e.g. \citealt{BandT,bs84,gh09}).  Orbits which conserve at least one isolating integral per 
degree of freedom are called {\it regular}.  It is the ubiquity of regular orbits in galaxies that permits the 
rich variety in their structure, mirroring their varied formation pathways \citep{schw79}.

The regular orbits in a static potential can be classified into families that conserve
qualitatively similar integrals of the motion, and therefore have similar morphologies.
Which orbital class a given star will occupy is determined by the available phase space
for different kinds of orbits in the potential, and its ICs - the phase space need not
be uniformly populated.  In a time-varying (or otherwise non-ideal) potential such as an 
ongoing merger, stars diffuse in the space of their conserved integrals, but not completely.  
If they do not cross boundaries between the orbital families, then their qualitative character 
may be preserved from the ICs.  Crossing between orbital boundaries can serve as a collective 
relaxation mechanism (e.g. \citealt{bh96}).  An intuitive grasp of the orbital classes is therefore
essential to understanding how dynamical systems evolve and relax.  

We begin with a brief overview of the types of regular orbits that are possible in various idealized 
potentials, leading up to a classification of the orbits in triaxial systems into families that 
conserve similar integrals.  For a more thorough and rigorous presentation we refer the reader 
to \citet{BandT}.

The simplest conceivable model is a spherical potential.  In this case the symmetry
about all three Cartesian axes implies conservation of the angular momentum
vector, so every orbit is confined to a plane.  The star oscillates in radius with frequency
$\omega_{r}$ while precessing in azimuth with frequency $\omega_{\theta}$.  If these two
frequencies are commensurate ($m \omega_{r} + n \omega_{\theta} = 0$ for some integers $m$ and $n$)
then the orbit closes on itself, as in a Kepler potential.  More generally, the orbits
form rosettes that eventually fill an annulus between the minimum and maximum of the radial
oscillations (pericenter and apocenter).

Very few galaxies are spherical, but many are consistent with axisymmetry.  In an axisymmetric
potential the angular momentum component about the symmetry axis, $L_{z}$, is conserved.  The direction
of $\vec{L}$ precesses about the $z$ axis as $L_{x}$ and $L_{y}$ vary.  The radial oscillations no
longer return the star to the exact same pericenter and apocenter every cycle, but are still bounded
between some $p=r_{min}$ and $a=r_{max}$.  

Two orbits with the same energy and $L_{z}$ can look quite
different from each other, ranging from orbits nearly confined to the $x-y$ plane, resembling eccentric
orbits in a thin disk, to puffed-up orbits nearly filling a spherical annulus over long times.
This suggests that the orbits are constrained by another integral in addition to $E$ and
$L_{z}$, related to how the energy is apportioned into vertical and radial motion.
Though it cannot be expressed analytically for a general axisymmetric potential, this third
integral ($I_{3}$) allows an assortment of stable axisymmetric systems, from thin disks
formed by quiescent accretion, to nearly spherical systems heated by persistent perturbations
or discrete encounters (e.g. \citealt{c60,c63a,hh64,saaf68,r82,BandT}).

\begin{figure}[htb]
\epsscale{1.17}
\plotone{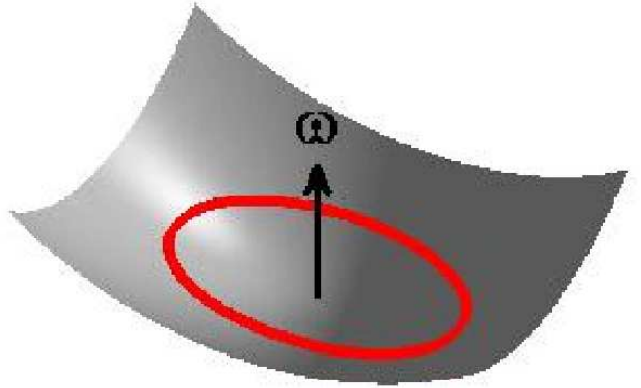}
\caption{{\bf Toy model of the box-tube boundary: Elliptical ring in a frictionless, concave trough.}
The ring is initially in equilibrium, with its long axis aligned with the long axis of
the trough.  If an angular velocity $\vec{\omega}$ is imparted to the ring, it must go up and over
a barrier to spin all the way around.  Above some critical $\omega_{c}$, the ring will make it
over the barrier and continue spinning; below $\omega_{c}$ it will just librate about the
equilibrium position.  This analogue owes to \citet{bs82}.}
\label{fig:ellipse}
\end{figure}

However some galaxies show clear evidence of nonaxisymmetric shapes (e.g. \citealt{fidz91}), 
and simulations of dissipationless violent relaxation in mergers or collapses generically 
produce triaxial remnants (e.g. \citealt{va82}).
There is a unique density distribution stratified on similar triaxial ellipsoids whose
potential is separable in ellipsoidal coordinates \citep{dzlb85}, known as the
``perfect ellipsoid''.  Its distribution function and integrals of motion can be expressed
analytically, and its orbital structure has been studied extensively \citep{stackel,
edd15,dzlb85,dz85,statler}.  The orbits in the perfect ellipsoid were classified into four major 
families by \citet{dz85}: short-axis tubes ($z-$tubes) which rotate about the
short ($z-$)axis of the potential, two classes of long-axis tubes ($x-$tubes) which revolve
about the long ($x-$)axis, and box orbits which behave like perturbed simple harmonic oscillators.  
Because it is analytic, this model has been widely used as a starting point for understanding 
the orbital structure of more general triaxial potentials.

Tube orbits resemble the orbits in axisymmetric potentials, and may be thought of loosely
as precessing ellipses driven by the bar-like potential of the triaxial ellipsoid
\citep{bs82}.  They conserve angular momentum-like integrals and therefore avoid the origin 
of the potential and the zero-velocity surface.  It can be shown that $y-$tube orbits rotating 
about the intermediate axis in the perfect ellipsoid are unstable to vertical perturbations 
\citep{hs79,dz85}, but both $x-$tube and $z-$tube orbits are allowed.  Observations of both 
major- and minor-axis rotation in some elliptical galaxies therefore strongly suggests that 
these systems are triaxial \citep{i77,sg79,b85,w88,fidz91}.  $x-$tube orbits are most prevalent 
in prolate potentials, and are populated by stars with large initial angular momenta about the 
long axis.  $z-$tubes are the dominant type of orbit in oblate systems.  Tube orbits oscillate 
in radius within some bounds $p$ and $a$, and one component of $\vec{L}$ never switches sign.  
Any net angular momentum of a triaxial system must be carried by the tube orbits, so only these 
orbits can retain information about a galaxy's initial sense of rotation.

Just as tube orbits may be thought of as precessing ellipses, box orbits may be regarded
loosely as axial orbits (or elongated ellipses) {\it librating} about the $x-$axis.
\citet{bs82} illustrated the transition between box and tube orbits using an intuitive toy 
model.  Imagine an elliptical ring lying in a concave, frictionless trough, with its long axis 
initially parallel to the long axis of the trough, as shown in Figure~\ref{fig:ellipse}.  
This configuration allows the ring to lie as low in the trough as possible.  Now imagine 
trying to spin the ring in the trough, by applying an impulsive kick of energy 
$\frac{1}{2} I \omega^{2}$ to one of its ends.  To spin all the way around the ring must go 
up and over a barrier, since the curvature of the trough along the perpendicular direction is 
greater.  There is some critical precession frequency $\omega_{c}$, above which the ring will 
get over the barrier.  Below $\omega_{c}$ it will just librate about the equilibrium configuration,
with no definite sense of rotation.  

The trough's different curvature along the two axes is analogous to triaxiality of a gravitational 
potential, and the librating mode is analogous to box orbits.  If the curvature is
the same along both directions perpendicular to the rotation axis (as in an axisymmetric
potential), then the energy barrier is zero.  Note also that the effective barrier in
a triaxial potential is greatest for a star rotating about the $y-$axis, since
in this case the difference in curvature along the two perpendicular directions is greatest, 
making $y-$tubes the most susceptible to instability.

Box orbits are prevalent in triaxial systems with shallow inner density profiles, and
conserve integrals similar to the energies of independent harmonic oscillations about 
each Cartesian axis.  Stars on box orbits have no definite sense of rotation, and can 
therefore pass arbitrarily close to both the origin (they are ``centrophilic'')
and the zero-velocity surface.  Over time, they densely fill a 3-dimensional box-like region 
centered on the origin \citep{dz85,statler,BandT}.  Powerful nuclear processes such as gas inflow, 
starbursts, and black hole growth are thought to be major drivers of galaxy evolution, 
and box orbits may be responsible for conveying information about the rapidly varying 
central potential to large radii.

\begin{figure}[htb]
\epsscale{1.17}
\plotone{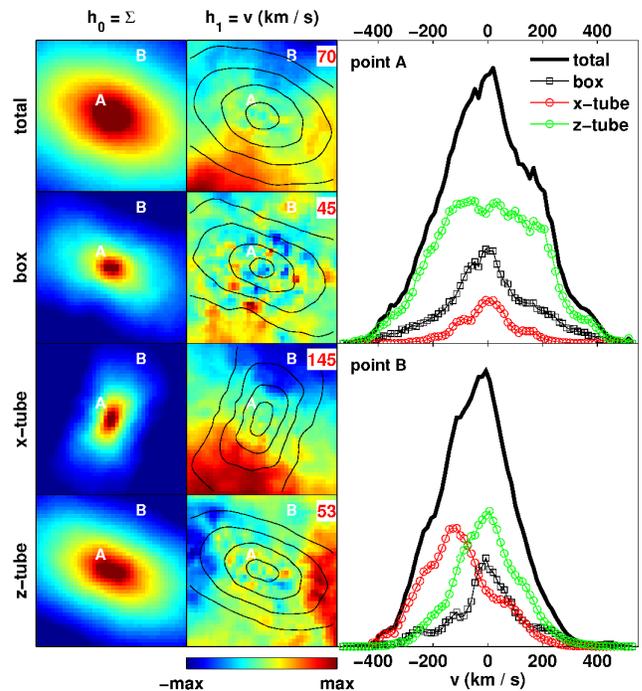}
\caption{{\bf Orbital composition of a typical merger remnant.}  Remnant $i$ (see Table 1), 15\% gas, viewed along a sight-line near the $y-$axis.  {\it First column:} Projected surface brightness, $\Sigma$, colored on a logarithmic scale.  All four maps use the same color scaling to show the relative contribution of each orbital class. {\it Second column:} LOS velocity $v$, with overplotted isophotes.  The size of the maps is 1$R_{e}$ $\times$ 1$R_{e}$, and the color scale runs from $-v_{max}$ to $v_{max}$, where $v_{max}$ is the maximum value of $|v|$ in any pixel.  The number in the upper right corner gives the value of $v_{max}$ in km/s.  The two right panels show the smoothed LOSVDs at the locations specified by the white letters on the $\Sigma$ and $v$ maps.  {\it Top:} LOSVD at point A, near the major axis.  {\it Bottom:} LOSVD at point B, near the minor axis.  The heavy black line is the full LOSVD, while the thin lines with markers break it down by orbital class.}
\label{fig:howAddsUp}
\end{figure}

Figure~\ref{fig:howAddsUp} shows an example of how the three orbital classes
contribute to the projected surface brightness and kinematics in one of our
simulated merger remnants.  The remnant is shown in projection along
the $y-$axis, which maximally separates the orbital classes in 2D space.  The boxes 
and $z-$tubes are elongated along the major axis, while the $x-$tubes are elongated 
along the minor axis of the projection.  All four surface brightness maps are 
plotted on the same color scale, to show the relative orbital populations at different 
locations in the sky plane.  Note that the isophotes of each individual orbital class 
appear boxier than the combined isophotes, which are nearly elliptical in shape.

The $x-$tube orbits are responsible for the minor-axis rotation in the remnant,
and the $z-$tubes produce the major-axis rotation.  There is a striking difference
in the amount of streaming of the two classes of tubes: the velocity scale
on the $x-$tube map is $\pm$145 km/s, while it is only $\pm$53 km/s on the $z-$tube
map.  The remnant therefore appears dominated by minor-axis rotation in the velocity
maps, even though $z-$tube orbits dominate its stellar mass.

Two local LOSVDs are shown on the right, at the locations indicated on the maps: 
one at a point near the center along the
major axis (point A) and one near the minor axis farther out (point B).
At point A the $z-$tubes have a flat-topped distribution with a high velocity
dispersion and negative kurtosis, indicative of canceling streams rotating
in opposite directions.  The box orbits are strongly peaked at $v=0$, with a
large excess in the tails giving the distribution a high positive kurtosis.
The combination of the $z-$tube and box orbits yields a combined LOSVD that
is closer to a Gaussian.  At point B the $x-$tube orbits make a larger relative
contribution.  Their LOSVD is peaked at a high negative velocity and skewed right,
indicative of a heated streaming population.  The combined distribution has
substantial net streaming motion and is slightly skewed in the direction
opposite the mean velocity.

An important idealization in the perfect ellipsoid model is that it asymptotes to a flat
core.  Most elliptical galaxies actually have central density cusps, with slopes ranging
from $d \ln{\rho} / dr \sim 0.5-2.5$ (e.g. \citealt{c93,f97,l05,l07,k09}), which may be 
thought of as perturbations to the perfect ellipsoid potential.  These perturbations
lead to trapping of box orbits by resonances (e.g. \citealt{BandT,cm77,st96,coll97,mv99,ty00}), 
$\vec{n} \cdot \vec{\omega} = 0$ for integer $\vec{n}$ (e.g. $z-$tube orbits are trapped by 
the 1:1 resonance between $\omega_{x}$ and $\omega_{y}$).  Stars in an 
island of phase space around a stable resonance librate around the resonant 
orbit; these trapped box orbits, called ``boxlets,''
generally avoid the origin (e.g. \citealt{specclass,mv99,fb01}).  As the perturbation grows 
larger (the cusp gets steeper), more phase space is occupied by resonant islands until they 
overlap, producing regions populated by ergodic orbits that eventually fill the entire portion 
of their 5D energy surface not occupied by regular orbits (e.g. \citealt{mv96,vm98,kansiop03,kal04}).    

For simplicity we do not distinguish between regular box orbits, boxlets, and ergodic
orbits in this paper, and will loosely use the term ``box'' to refer to any orbit with 
no definite sense of rotation.  This will be sufficient for the global comparison with 
observations desired in this study.  Spectral analysis of selected orbits \citep{thesis,gaspard} 
reveals that the orbits classified as boxes are generically centrophilic (at least in the
inner parts) and stochastic, but they remain fairly localized in phase space over a Hubble 
time.  We defer a detailed study of the nature of the box orbits in the remnants to future 
work.

The effect of an inner cusp on the orbital populations can be understood by means of
a simple experiment - adding a central point mass to a flat-cored, triaxial potential
with a large population of box orbits.  This ``experiment'' has been studied extensively,
since it has direct astrophysical relevance to black holes in the nuclei of galaxies
(e.g. \citealt{norm85,gb85,quin95,sig95,mv97,mq98,vm98,pm01,holl02,kal04}).  Since 
box orbits pass arbitrarily close to the origin, they are deflected by
the point mass at pericenter passage, causing the orbits to diffuse within box phase space.
The velocity changes are primarily along the axis of approach \citep{chandradf}, the $x-$axis for 
box orbits librating about the long axial orbit, so the angular momentum diffusion is mostly in the
$y-z$ plane.  Eventually the star may wander into $z-$tube phase space (there is no space
for $y-$tubes since they are unstable).  Since box orbits also strike the zero-velocity
surface, they convey information about their pericentric evolution to large radii.  

As box orbits diffuse across the $z-$tube boundary, the shape of the potential also becomes more
oblate, shrinking the phase space available for boxes and expanding that for $z-$tubes
(e.g. \citealt{kal04}).  A point mass as small as 2-3\% of the total mass can seed this transformation
in a flat-cored, triaxial system \citep{gb85,mv97,mq98,vm98}.  When the 
central mass concentration (CMC) is not an ideal point mass,
the diffusion timescale varies with the degree of central concentration - in an $r^{-1}$
cusp the timescales are typically longer than the lifetime of a galaxy, while in an
$r^{-2}$ cusp they are short \citep{mv96,mf96,fm97,holl01}.  In hierarchical structure formation, 
there is an ongoing exchange between processes that induce triaxiality (e.g. violent relaxation in
mergers) and gas inflows that deepen the central potential well (e.g. \citealt{d94}).  This
process therefore undoubtedly plays an important role in galaxy evolution.       

\subsection{Gas-rich mergers}

In current semi-analytic models, $\sim 70$\% of the stellar mass in present-day
ellipticals and classical bulges assembles through major mergers \citep{hop09a,hop09d,fm08,cw09}.
An elliptical galaxy-sized halo has on average undergone $\sim 1$ major merger since $z \sim 2-3$, the
epoch during which most of its stellar mass formed \citep{k96,d06,hop08a,hop09a,s08}.  The last 
major merger was typically between two spiral galaxies, with gas fractions ranging from $\sim 10\%$ for 
systems with stellar masses around $3 \times 10^{11} M_{\odot}$, to $\sim 50\%$ for 
$10^{10} M_{\odot}$ systems (\citealt{s09a,erb06} and references therein).

The hypothesis that elliptical galaxies form through mergers between spirals \citep{tt72} actually 
preceded the acceptance of the concordance cosmology by about two decades, based on the properties 
of the galaxies themselves.  Gas-rich tidal tails, and rings and shells indicative
of the recent disruption of a dynamically cold system, often surround galaxies
otherwise resembling ordinary giant ellipticals \citep{a66,vd05}.  Early simulations of
mergers between disk galaxies could explain a wide variety of the properties of observed
ellipticals, including their “$r^{1/4}$” law density profiles \citep{va82,mg84,hernquist92,naabtruj06},
slow rotation and anisotropic velocity distributions \citep{op73,ab78,w78,g81}, 
flat rotation curves \citep{w78,fs82,e82}, fine structure \citep{hs92},
and apparently triaxial shapes \citep{g81,barnes88,fidz91}.
A simple counting argument based on the numbers of observed interacting pairs and elliptical
galaxies in the local universe made the prospect that these pairs turn into ellipticals 
quite plausible \citep{toom77}.

It was apparent that dissipation must play a large role in mergers long before
hydrodynamic simulations with realistic gas fractions became feasible 
\citep{nw83,hernquist89,nieto91,bh91gas,lutz91,az92,hsh93}.  The measured phase space densities at the 
centers of elliptical galaxies far exceed the maximum densities in observed spirals, 
implying a violation of Liouville's theorem unless the initial disks contain $\gtrsim$25-30\%
of their mass in gas \citep{c86,hsh93,r06fp}.  Dynamically cold components, such
as embedded disks and KDCs, are often observed in elliptical galaxies 
(e.g. \citealt{fi88,js88,hb91kdc,rw90,scorz95,jess07}).  The high specific frequencies of GCs in ellipticals
relative to spirals imply that mergers must trigger the formation of many new
clusters from the available gas \citep{az92,gcmf}.

More recent simulations have shown that mergers with $\sim 30$\% gas produce remnants that fall on the 
observed fundamental plane scaling relation \citep{dd87,d87,gonz03,r06fp,hop08c},
and that the remnants of 1:1 mergers between 40\% gas disks match the 1D kinematic properties of observed
ellipticals far better than dissipationless merger remnants \citep{tjkin}. 
2D kinematic maps of gas-rich merger remnants display many of the intriguing features seen in real galaxies, 
including misaligned rotation, central velocity dispersion dips, counter-rotating disks, and KDCs 
\citep{jess07,barnes92,hb91kdc}.  The shapes of the LOSVDs of simulated gas-rich merger remnants display 
the same trends as ellipticals in the SAURON sample \citep{b94,bb00,nb01,bour05b,gg06,naab06a,h3h4}, and 
they occupy essentially the same part of the anisotropy-ellipticity ($\delta-\epsilon$) plane as the SAURON 
galaxies (with the notable exception of some giant slow rotators that are very round, anisotropic, and 
featureless; \citealt{bur08}).  These results motivate studies of the intrinsic orbital structure of gas-rich 
merger remnants, to see to what extent the 3D distribution functions of elliptical galaxies can indeed be 
explained with binary mergers, and shed light on how the observable features arise physically.

To our knowledge, \citet{barnes92} was the first to apply orbital analysis in the style of \citet{dz85,statler}
to simulated merger remnants.  He ran a series of dissipationless disk galaxy mergers with varying disk orientations
and impact parameters, and classified the stellar orbits in the remnants based on the sign changes in their angular 
momentum.  He found a wide variety in the remnant shapes and orbital structure, depending on the encounter 
parameters.  Some of the remnants displayed substantial orientation twists (see also \citealt{g83}).  The mergers 
often produced both $x-$tube and $z-$tube populations with significant net rotation, resulting in large kinematic 
misalignments.  Since the rotation of the majority of observed ellipticals is well-aligned with the major axis, he 
argued that dissipationless disk mergers cannot be the generic way to form elliptical galaxies.

\citet{bh96} classified the stellar orbits in mergers with 10\% gas, and found that even this small gas component
has a dramatic effect on the remnant shapes and orbital structure.  Gravitational torques during the merger drain 
much of the gas of its angular momentum, causing it to collapse inward and form a dense CMC 
\citep{hernquist89,bh91gas,mh94a,mh94b,mh96,hop09c}, essentially a point mass to stars at $1R_{e}$.  The CMC destabilizes box 
orbits, leading to a global transformation of the remnant to a more oblate shape.  This result is not surprising 
in light of the studies by e.g. \citet{gb85,mq98}, showing that a central point mass with just 2\% of the total 
mass can globally transform the structure of a triaxial system.

\citet{jess05,naab06a} performed a detailed study of the orbital structure of a large sample of equal and unequal mass merger
remnants with 0 and 10\% gas \citep{nb03}, with an emphasis on relating the intrinsic structure to photometric and 1D kinematic
observables.  They studied the relation between the orbital structure and isophotal shapes in further detail in \citet{jess08}. 
Using spectral classification \citep{specclass}, they found that box orbits typically dominate the inner parts of
the dissipationless remnants, while $x-$tubes and $z-$tubes become dominant at larger radii.  The box to $z-$tube ratio was the
primary determinant of kinematic properties such as the location of the remnants in the $h_{3,4}-v / \sigma$ planes.   
When a gas component was added, the box population was highly suppressed and the remnants became $z-$tube dominated and oblate.
The shape of the $z-$tube orbits in the more axisymmetric dissipative remnants made the isophotes less boxy.  The 1:1 merger
remnants were slowly rotating, while the 3:1 remnants were found to be rapidly rotating and disky.  They concluded that observed
rapidly rotating ellipticals could form from dissipative 3:1 mergers, but boxy, slowly rotating systems \citep{kb96} could not 
have formed through dissipative disk mergers.  

In this work we analyze the orbital structure of 1:1 merger remnants in simulations including 
SF and feedback, enabling us to consider the high gas fractions characteristic of 
spiral galaxies at $z \sim 2$ \citep{s09a,gasz15}.  At fixed $f_{gas}$ the effect on the remnant
structure may be diminished by including SF and the dissipational features may become more
spatially extended, since the gas can be converted to collisionless material
early on in the merger and undergo subsequent violent relaxation.
We quantify the variation of the orbital structure and intrinsic shape with gas fraction for $f_{gas}$
ranging from 0 to 40\%, and discuss the physical mechanisms that may be driving the orbital
transformations, with direct comparisons against dynamical models of observed systems in mind.  
We relate the orbital structure of the remnants to their appearance in 2D kinematic maps on $1R_{e}$ 
and $3R_{e}$ scales, and show that a wide range of different kinematic structures can be accounted for 
simply by varying $f_{gas}$ in 1:1 disk mergers.

\subsection{Outline}

In \S2 we describe our merger simulations and remnant analysis methods, and in \S3 we present our results.  
The intrinsic structure of our eight dissipationless remnants is presented in \S3.1, as a baseline 
for understanding the effect of adding a gas component.  We show how the orbital structure and shape of the 
remnants varies with $f_{gas}$ in \S3.2, and discuss the KDCs and embedded disks that arise in the gas-rich
remnants in \S3.3.  In \S3.4 we briefly describe the structure of the remnants beyond $\sim 1R_{e}$, deferring
a more detailed and quantitative presentation to \citet{hoff09b}.  In \S4 we summarize our results and conclude.
The radial profiles of the orbital structure, intrinsic shape, and orientation of all 48 dissipative remnants
are presented in the Appendix. 

\section{Simulations and methods}

\subsection{Simulations}

\begin{table}
\vspace{0.3in}
\caption{Merger orbits}
\vspace{-0.1in}
\begin{center}
\renewcommand{\arraystretch}{1.4}
\begin{tabular}{c c c c c}
{\bf Orbit} & $\theta_{1}$ & $\phi_{1}$ & $\theta_{2}$ & $\phi_{2}$ \\ \hline
$i$ & 0 & 0 & 71 & 30 \\
$j$ & -109 & 90 & 71 & 90 \\
$k$ & -109 & -30 & 71 & -30 \\
$l$ & -109 & 30 & 180 & 0 \\
$m$ & 0 & 0 & 71 & 90 \\
$n$ & -109 & -30 & 71 & 30 \\
$o$ & -109 & 30 & 71 & -30 \\
$p$ & -109 & 90 & 180 & 0 \\ \hline
\end{tabular}
\end{center}
\label{mergerorb}
\end{table}

Our galaxy merger simulations were performed using the publicly available TreeSPH 
(Smoothed Particle Hydrodynamics) code Gadget-2 \citep{gadget2,gadget}, which uses an advanced 
formulation of SPH that explicitly conserves both energy and entropy when appropriate 
\citep{gaden}.  In addition to the standard features, our version of the software includes 
sub-resolution prescriptions for radiative cooling \citep{katz96,dave99}, 
SF, and feedback from 
supernovae and AGN, as described in detail in \citet{gadfeedbak}.

SF and supernova feedback are treated using the multi-phase prescription 
of \citet{gadsf} and \citet{gadfeedbak}. The interstellar medium (ISM) consists of a cold cloud 
phase in which stars form, and a hot phase that provides pressure support for the disk.  
The density dependence of the SF rate is calibrated to match the observed 
Schmidt-Kennicutt Law \citep{schmidt,kenn}.  Radiative cooling drives gas transfer from the hot phase to 
the cold phase, while heating from supernovae feeds gas back into the hot phase.  The 
parameter $q_{EOS}$ smoothly varies the effective equation of state (EOS) between isothermal 
($q_{EOS}=0$) and the pure multi-phase model ($q_{EOS}=1$); higher values of $q_{EOS}$ 
correspond to a stiffer EOS and permit stable disks at higher gas fractions.  In all of the
simulations in this paper we used $q_{EOS}=0.25$.

The simulations also include sink particles representing black holes, which accrete gas at 
a rate $\dot{M}_{BH} = \min (\dot{M}_{Bondi},\dot{M}_{Edd})$, where 
$\dot{M}_{Bondi} \propto M_{BH}^{2} \rho_{gas} / \sqrt{c_{s}^{2}+v^{2}}$ is a rate based on 
the Bondi-Hoyle-Lyttleton formula \citep{bondi52,bh44,hl39}, and $\dot{M}_{Edd}$ is the Eddington rate. 
A fraction $\epsilon_{f} \epsilon_{r}$ of the accretion energy is released 
thermally and isotropically into the surrounding gas, where $\epsilon_{r}=0.1$ is the 
assumed radiative efficiency, and $\epsilon_{f}=0.05$ is the fraction of the radiated 
luminosity that can couple thermally to the gas.

The gravitational softening length was $\epsilon = 140$ pc for the stellar and gas components 
in our simulations, and $\epsilon = 290$ pc for the halo component. We decreased the 
parameter $\eta$ controlling the Gadget-2 timestep criterion 
($\Delta t = \sqrt{2 \eta \epsilon / |\vec{a}|}$, where $\vec{a}$ is the particle 
acceleration) from the default value of 0.025 to 0.0025, to ensure that the steep central 
cusp that formed in the gas-rich simulations was preserved for several Gyrs after the merger.

We set up our merger ICs using standard techniques presented in 
\citet{gadfeedbak}, which have been employed in a variety of previous 
applications (e.g. \citealt{dsh05,naab06a,hop06,r06fp,tjkin,johan09,hop09b}).  The galaxy models 
consisted of exponential disks embedded in dark matter halos; stellar bulges were not included 
in this study.  The total mass 
was specified through the virial velocity, $v_{200} = 160$ km/s for an approximately Milky 
Way-sized galaxy, by $M_{tot} = v_{200}^{3}/(10GH_{0})$.  The halo was modeled as a 
\citet{hernquist90} profile with spin parameter $\lambda = 0.033$, and scale radius chosen
to match an NFW profile with concentration $c = 0.9$ in the inner parts.

The disk mass was set to $M_{d} = f_{d}M_{tot}$, with $f_{d} = 0.041$. Its scale length
of $r_{d} = 3.9$ kpc was determined by the requirement that it be centrifugally supported
with the same specific angular momentum as the halo, $J_{d} = f_{d}J_{tot}$.  The stellar 
component of the disk was assigned a radially constant vertical scale height of 0.2$r_{d}$, 
while the vertical structure of the gas component was set by the requirement of hydrostatic 
equilibrium.  The disk models were realized with 80000 equal-mass particles, a fraction 
$1-f_{gas}$ of them in collisionless stars and the other $f_{gas}$ in SPH particles.
The halo was comprised of 120000 collisionless dark matter particles. 

For each simulation two identical disk galaxies were constructed following this procedure, 
and placed on a parabolic orbit with an impact parameter of 7.1 kpc.  At their initial 
separation of 140 kpc, the center-of-mass trajectories were well-approximated by point mass
orbits.  At least in the inner parts, the final structure is not highly sensitive to our 
choice of small impact parameter, since the most bound particles (the stars) lose most of 
their orbital angular momentum to the halo by the time the cores merge even in wide
encounters \citep{barnes98,tjkin}.  The effect of the merger impact parameter on the stellar 
halo structure will be explored in future work \citep{hoff09b}.

The orientation of the disks relative to the merger ($x-y$) plane was parametrized by the two
angles $\theta$ and $\phi$, the polar angle of the spin vector relative to the $z-$axis and 
its azimuthal angle relative to the axis of approach, as shown in Figure 6 of \citet{tt72}. We 
used the scheme proposed by \citet{barnes92} to sample the space of possible disk 
orientations in a relatively unbiased way, including prograde, retrograde, and polar orbits.
The spin of the first disk coincides with each of the four symmetry axes of a regular 
tetrahedron pointing upward, while that of the second disk points along the axes of the
corresponding downward-pointing tetrahedron. The resulting eight merger orbits are listed in 
Table 1. This set of encounter orbits has been used in a number of previous studies 
(e.g. \citealt{barnes92,nb03,naab06a,tjkin}), which may be used as a reference point for 
comparison with our results where appropriate.
The eight encounters listed in Table 1 were repeated at seven different gas fractions,
$f_{gas}$ $=$ 0, 5, 10, 15, 20, 30, and 40\%, for a total of 56 merger simulations.

Once on their orbits, the galaxies reach their first pericenter passage at 
$t \approx 0.35$ Gyrs, at which point their morphologies become strongly distorted by tidal 
forces from the other galaxy, and the stars and gas each form a bar.
The stellar bar slightly trails the gas bar, torquing back on the gas and draining its
angular momentum so that the gas flows inward and undergoes a burst of star SF 
\citep{mh94a,bh96,escala04,hop09b}.  At $t \approx 1.8$ Gyrs the galaxies reach second passage, 
and their cores merge shortly thereafter.  The strong and persistent gravitational torques during the 
second passage and final merger typically induce a stronger central starburst than at first 
passage.  By about 0.5 Gyrs after the merger the size, shape, and velocity dispersion of
the remnant (measured within $\sim 1R_{e}$) reach a steady state \citep{tjkin}, and the system
may be considered relaxed.

\subsection{Remnant analysis}

We freeze the potential at $t=4.3$ Gyrs (about 2.5 Gyrs after the merger is complete),
and represent it using the bi-orthogonal ``designer'' basis expansion \citep{cb72,cb73,BandT} 
presented in \citet{scf},
\begin{eqnarray}
\rho(r,\theta,\phi) & = & \sum_{nlm} A_{nlm} \rho_{nl}(r) Y_{lm}(\theta,\phi) \\
\Phi(r,\theta,\phi) & = & \sum_{nlm} B_{nlm} \Phi_{nl}(r) Y_{lm}(\theta,\phi).
\end{eqnarray} 
The terms in this expansion individually satisfy Poisson's equation, and the lowest-order 
term is a \citet{hernquist90} profile, which should be a good first approximation to the density
profile of a remnant resembling an elliptical galaxy.  Before computing the expansion coefficients
we symmetrize the particle distribution by reflecting all of the positions and velocities about
the origin, effectively increasing the particle statistics by a factor of two and reducing global 
fluctuations.  The orbit of each stellar particle is then integrated through $\sim 300$ radial 
turning points (peri/apocenter passages) in the static potential using a Bulirsch-Stoer integrator 
\citep{bs,numrec}.  This technique allows all of the stellar orbits to be efficiently followed for 
$\sim$150 dynamical times, including halo orbits for which this corresponds to many Hubble times.  
It also lessens spurious relaxation and simplifies the orbital analysis by fixing the principle 
axes of the system.

Each stellar orbit was classified as a box, $x-$tube, or $z-$tube orbit using a simple
algorithm based on the sign changes in the star's angular momentum, introduced by 
\citet{barnes92}.  At each timestep we computed the angular momentum vector $\vec{j}$, and 
checked whether each component of $\vec{j}$ had changed sign since the last step.  At the 
end of the integration we constructed the vector $\vec{k}$ such that $k_{i}=1$ if $j_{i}$ 
never changed sign, and $k_{i}=0$ otherwise.  The orbit was assigned the classification code 
$C = k_{x}+2k_{y}+4k_{z}$, which is 0 for a box orbit, 1 for an $x-$tube orbit, and 4
for a $z-$tube orbit.  For a perfect triaxial ellipsoid, other values of $C$ are unphysical.

This classification scheme depends on a correct specification of the system orientation, 
and many of the remnants have intrinsic orientation twists.  We therefore diagonalized the
inertia tensor, in the form $I_{ij} = \sum x_{i}x_{j}$ (e.g. \citealt{diag1}), in $\sim 50$ cumulative energy bins, 
and used the local orientation based on the star's energy for the classification.  The 
eigenvectors of $I_{ij}$ are the principal axes of the remnant's figure, and the square roots 
of the eigenvalues give the relative scale lengths along the three principal axes, $a$, $b$, 
and $c$.  All particles - gas, stars, and dark matter - were included in $I_{ij}$, since a
star's orbit is determined by the sum of all three components.  Note that since we used the
local orientation for the classification, e.g. the position angle of the $z-$tube rotation 
may appear to vary with galactocentric radius in the remnants with significant orientation 
twists.

Only $\sim$1\% of the stellar orbits were not assigned $C$ values of 0, 1, or 4, except in
rare instances of sudden orientation twists or locations where the potential was very nearly
spherical.  The results of our simple classification scheme were consistent with spectral 
classification \citep{specclass,bs82,bs84,thesis}, and we determined that 
the simpler algorithm was better suited for the noisy potentials and gross analysis desired 
in this work.  However note that this algorithm does not distinguish 
between boxes, resonant boxlets, and ergodic orbits.

\begin{figure*}[tb]
\epsscale{1.2}
\plotone{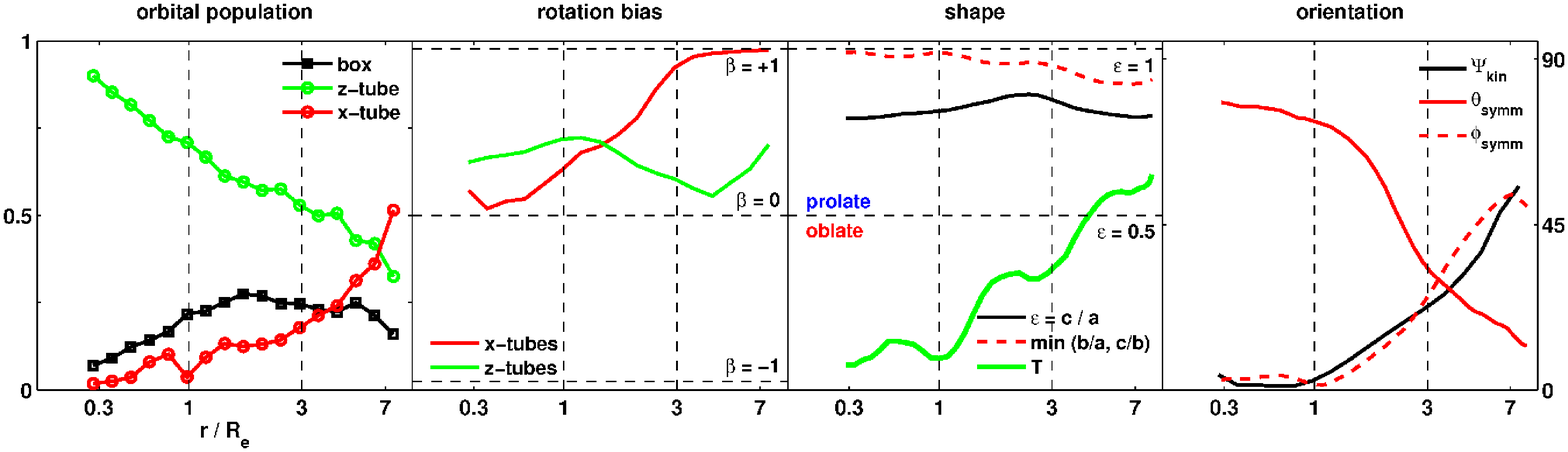}
\caption{{\bf Radial profiles of the orbital structure and intrinsic shape of remnant $i$, 40\% gas.}  {\it Left:} Fraction of the stellar mass in each of the three major orbital classes vs. percentile in binding energy.  The horizontal axis is labeled by the mean radius (in units of $R_{e}$) of the stars in the current energy bin (see text for details).   {\it Left center:} Rotation bias, $\beta$ $=$ $(N_{+}-N_{-}) / (N_{+}+N_{-})$, of the $x-$tube and $z-$tube orbits.  $\beta = 0$ corresponds to perfect cancellation of the rotation, while $\beta = \pm 1$ means that all of the stars are streaming in the same direction.  We choose the overall sign so that $\beta$ is positive at the 40th percentile in binding energy ($1R_{e}$).   {\it Right center:} Intrinsic shapes of the remnants.  The heavy green line is the triaxiality parameter, $T$, for all particles in the current energy bin and more bound.  The black line is the intrinsic ellipticity $\epsilon = c / a$, and the red dashed line is the smaller of the other two axis ratios.  The horizontal line at $T = 0.5$ marks the boundary between oblate and prolate shapes.  Both stars and DM particles are included in the evaluation of the intrinsic shapes.  {\it Right:} Intrinsic orientations ($\theta_{symm}, \phi_{symm}$) and kinematic misalignments ($\Psi_{kin}$), in degrees. The red solid and dashed lines are the polar ($\theta_{symm}$) and azimuthal ($\phi_{symm}$) angles of the symmetry axis, in this case the short axis since the remnant is oblate.  In general we will plot $\theta, \phi$ of the short axis in red if the remnant is oblate ($T<0.5$) at $1R_{e}$, and $\theta, \phi$ of the long axis in blue if the remnant is prolate ($T>0.5$) at $1R_{e}$.  $\theta$ and $\phi$ are evaluated in a coordinate system where the disks' initial orbital angular momentum points in the $\hat{z}$ direction, and the galaxies initially approach one another along the $x$-axis.  To evaluate $\Psi_{kin}$ (equation 3; black line), we return to the local principal axes coordinate system.  In all panels, the $1R_{e}$ scale probed by SAURON and the $3R_{e}$ scale typical of SKiMS data are marked with vertical lines.}
\vspace{0.1in}
\label{fig:orbexample}
\end{figure*}

For each remnant, we present radial profiles of the orbital structure and 
intrinsic shape as shown for one example in Figure~\ref{fig:orbexample}.  First we order 
the stellar particles by energy, and divide them into $\sim 20$ energy bins with equal numbers of 
particles.  The mass fraction of stars in box, $x-$tube, and $z-$tube orbits is computed for 
each energy bin, and plotted vs. percentile in binding energy (by mass), as in the left-hand panel of 
the figure.  For ease of comparison with observational results, we also compute the mean 
galactocentric radius of the particles in each energy bin, and label
the bins corresponding to 0.3, 1, 3, and the largest integer number of $R_{e}$.

For understanding kinematic maps, it is useful to quantify not only the populations of the
major orbital classes, but also their degree of streaming.  For this purpose we plot profiles of
\begin{equation} 
\beta=(N_{+}-N_{-})/(N_{+}+N_{-})
\end{equation} 
for the $x-$tube and $z-$tube orbits, as shown in the left center panel.  Here $N_{+}$ and 
$N_{-}$ are the numbers of particles rotating in either direction; $\beta = 0$ corresponds 
to perfect cancellation of the rotation, while $\beta = \pm 1$ means that all of the stars 
are streaming in the same direction.  The sign convention was chosen by requiring $\beta$
to be positive at the 40th percentile in binding energy ($1R_{e}$).

The intrinsic shape and orientation vector are computed by diagonalizing
the inertia tensor as described previously.  We plot the maximum and minimum
axis ratios, $\epsilon = c/a$ and $r_{min} = \min (b/a,c/b)$, and the triaxiality parameter, 
\begin{equation}
T=(a^{2}-b^{2})/(a^{2}-c^{2}),
\end{equation}
in the right center panel of Figure~\ref{fig:orbexample}.
$T$ ranges from 0 for a perfectly oblate spheroid to 1 for a prolate spheroid.  In the 
right-hand panel we plot the polar and azimuthal angles ($\theta_{symm}$ and $\phi_{symm}$) of 
the symmetry axis - the long axis if $T > 0.5$ within $1R_{e}$, or the short axis if $T < 0.5$.  If 
the short axis orientation is plotted then we color the lines red; otherwise we color them blue.  
The shape of a remnant that is oblate within $1R_{e}$ may become prolate farther out; in this case 
we still plot the orientation of the short axis at all radii, which does not make the angles ill-defined 
since the remnants are never too close to axisymmetric at large radii.  Note that $\theta_{symm}$ and
$\phi_{symm}$ represent angles between a vector and an ${\it axis}$ (whose direction is defined only
up to a sign), and therefore range from 0 to $90^{\circ}$.  When plotting the orientation profiles,
we use a fixed coordinate system where the $z-$axis points along the disks' initial orbital angular
momentum vector, and the $x-$axis is along the initial direction of approach.  

On the same axes we plot the intrinsic kinematic misalignment,
\begin{equation}
\Psi_{kin}= \arctan |j_{z}/j_{x}|, 
\end{equation}
or the angle between the short axis and the net angular momentum vector.  To compute $\Psi_{kin}$ we
revert back to the local principal axes system, so that $\Psi_{kin} = 0$ always corresponds to rotation
about the intrinsic short axis. 

To place our work in an observational context, it is also useful to show how the intrinsic
orbital structure imprints itself on SAURON-like 2D kinematic maps.  We constructed histograms of 
the LOS velocity in 40 $\times$ 40 spatial bins within $1R_{e}$, using 80 velocity bins within 
$\pm 3.5\sigma$.  For each LOSVD we performed a least-squares fit to the 5-parameter function
\begin{equation}
F(y) = A e^{-w^{2}/2} [1 + h_{3} H_{3} (w) + h_{4} H_{4} (w)],
\end{equation}
where $w=(y-v)/\sigma$, and $v$ and $\sigma$ represent the mean velocity and velocity
dispersion.  Here $H_{3}$ and $H_{4}$ are Gauss-Hermite (GH) polynomials, and the
parameters $h_{3}$ and $h_{4}$ measure the skewness and kurtosis of the distribution \citep{vf93,g93,BandM}.
The Gadget particles were smoothed over a radius $h_{smooth}=\max(h_{see},h_{ngb})$, 
where $h_{see} = 150$ pc corresponds to a seeing of 1.5'' at 20Mpc, and $h_{ngb}$ is 1.7 
times the distance to the 128th nearest neighbor.

We focus primarily on the velocity fields in this work.  Note that $v$, as derived from the fit of 
equation 4, deviates systematically from the true mean of the distribution for nonzero $h_{3}$ and $h_{4}$ 
(see \citealt{vf93} for the relevant correction terms).  However we follow the convention of previous authors 
(e.g. \citealt{b94}) and simply use $v$ from the GH fit to represent the velocity in this paper. 

When displaying all of the remnants at a given $f_{gas}$, we choose the viewing angles randomly (from
an isotropic distribution) to give a fair representation of their appearance in kinematic maps.  We
label each map with the LOS angles $\theta$ and $\phi$, where $\theta$ is the polar angle between the 
LOS and the $z-$axis, and $\phi$ is the azimuthal angle between the LOS and the $y-$axis.  The apparent
major-axis rotation is maximized when $\theta = 90^{\circ}$, and the minor-axis rotation is maximal at 
$\phi=0$ for a given $\theta$. 

\section{Results and discussion}

\subsection{The dissipationless remnants}

\begin{figure*}[p]
\epsscale{1.2}
\plotone{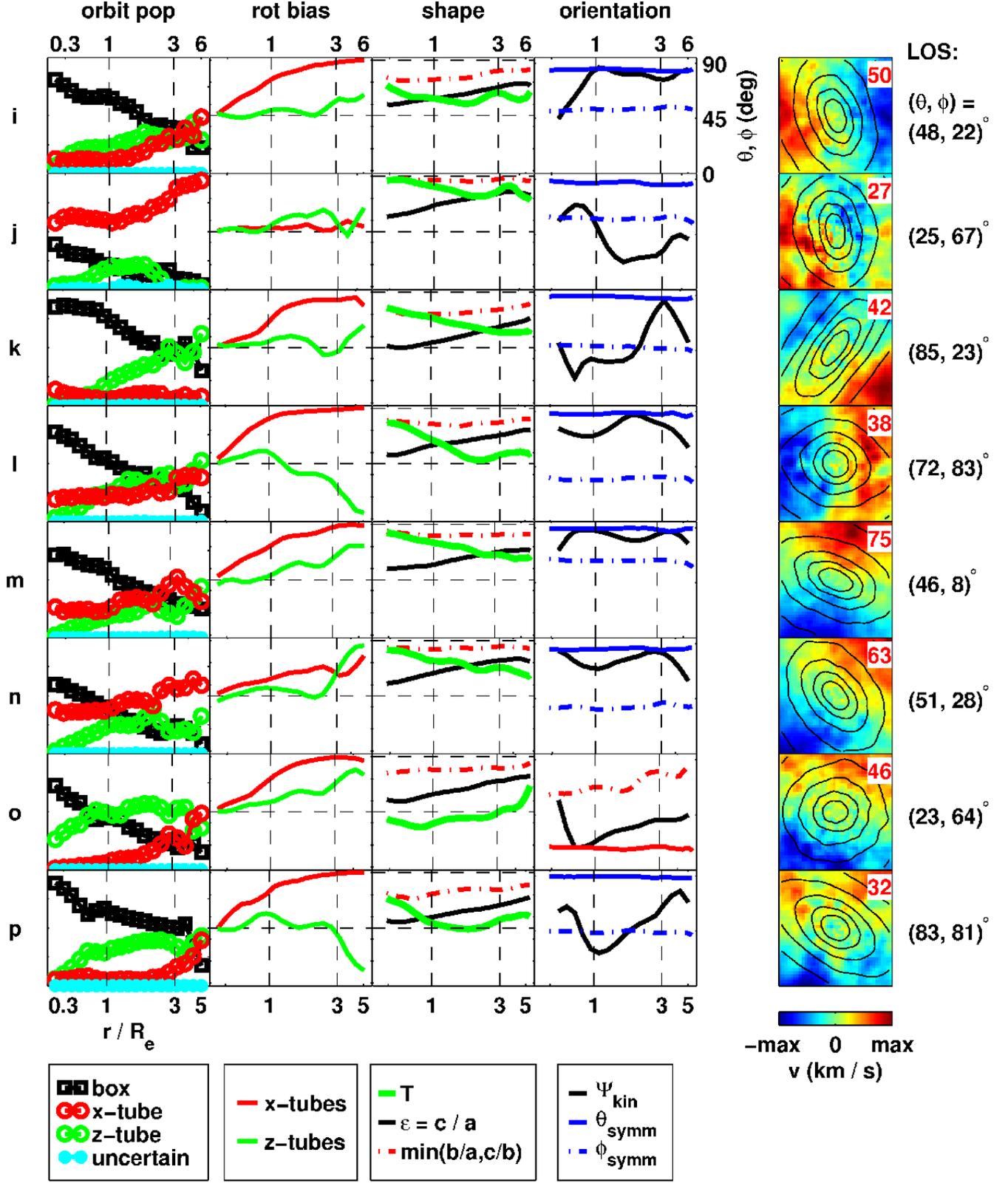}
\caption{{\bf Intrinsic structure of the dissipationless remnants.}  The first four columns follow the format of Figure~\ref{fig:orbexample}.  The merger orbit (see Table 1) is specified on the left for each row.  In the right-hand column we show a $1R_{e}$ velocity map of the remnant, viewed along a randomly (isotropically) chosen LOS.  The two angles specifying the LOS are indicated by the labels on the right.  $\theta$ is the polar angle between the LOS and the short axis (this axis was chosen as the reference throughout the paper since most of the remnants with gas are oblate), and $\phi$ is the azimuthal angle between the LOS and the intermediate axis.  The major-axis rotation is maximal at $\theta = 90^{\circ}$, and the minor-axis rotation is maximized at $\phi=0$ for fixed $\theta$.  The velocity scale runs from $-v_{max}$ to $v_{max}$, where $v_{max}$ is the maximum value of $|v|$.  The value of $v_{max}$ (in km/s) is indicated by the label in the upper right corner of each map.}
\label{fig:rems0}
\end{figure*}

\begin{figure*}[p]
\epsscale{1.2}
\plotone{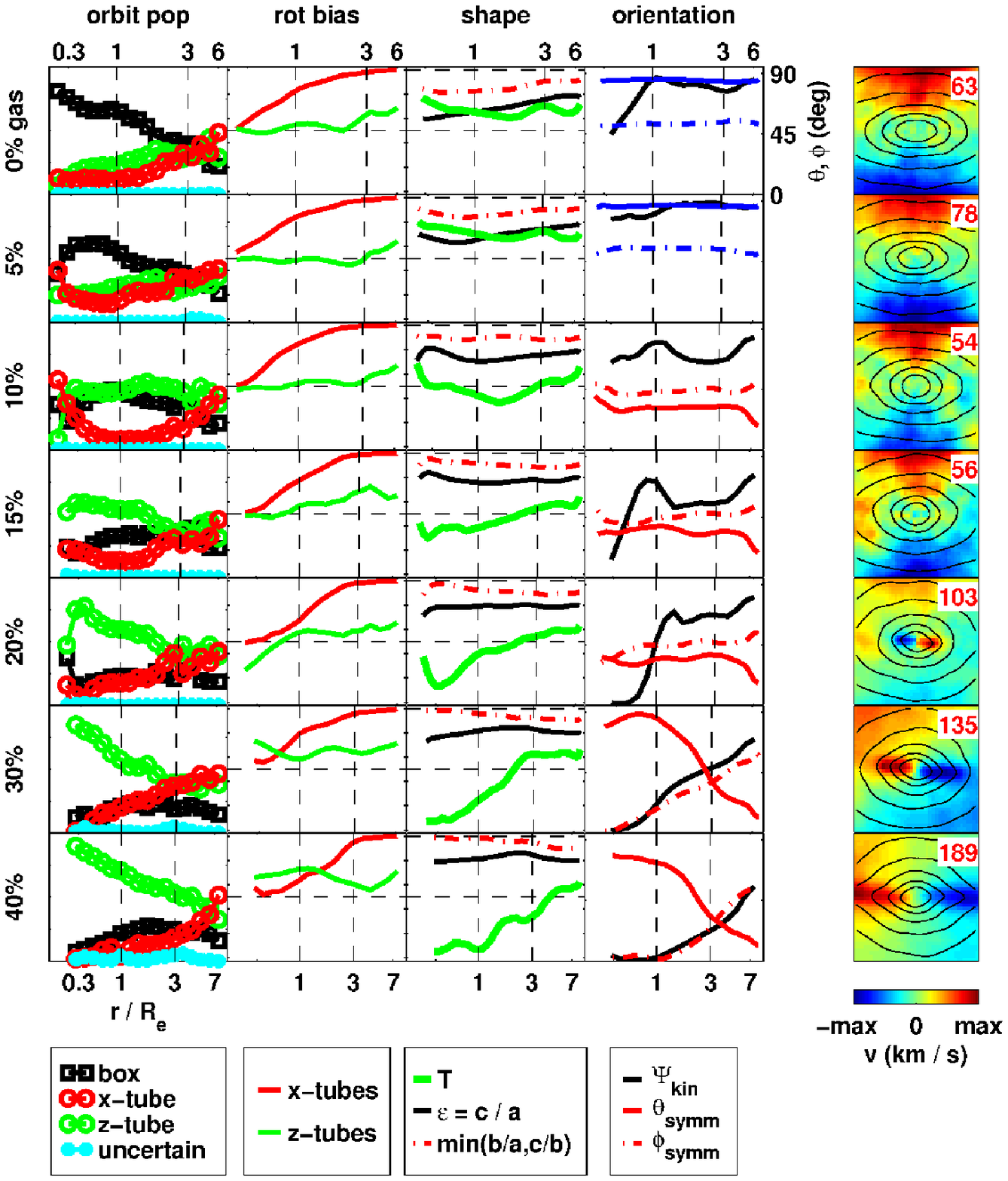}
\caption{{\bf Evolution of the intrinsic structure of remnant $i$ with gas fraction.}  The format is the same as in Figure~\ref{fig:rems0}, except that now each row corresponds to a different $f_{gas}$.  In this figure all of the velocity maps are shown in projection along the intermediate axis, $(\theta, \phi)=(0, 0)$, to allow the major- and minor-axis rotation in the remnants at different $f_{gas}$ to be compared directly.}
\vspace{0.1in}
\label{fig:igfs}
\end{figure*}
\suppressfloats

The structure of the dissipationless remnants is presented in Figure~\ref{fig:rems0}.  There is 
substantial variation in the orbital distribution and shape over the eight merger orbits - for
instance remnant $j$ is uniformly prolate and dominated by $x-$tube orbits, while remnant $o$ is 
oblate-triaxial and composed primarily of box orbits and $z-$tubes.  Overall the remnants are 
dominated by box orbits in their inner parts ($r \lesssim 1.5R_{e}$), and tube orbits in their 
outskirts.  This is not surprising since the \citet{hernquist90} profile, which generically
provides a good fit to the density profiles of dissipationless merger remnants, has a shallow
inner cusp ($\rho \propto r^{-1}$) and much steeper outer profile ($\rho \propto r^{-4}$)
\footnote[1]{A star in what we are calling the ``outer parts'' ($\sim 3-6R_{e}$) still
occupies the {\it inner} DM halo (which also follows roughly a \citealt{hernquist90} profile but with
a much larger scale radius).  The total density (stars $+$ DM) at these radii scales roughly like $r^{-2}$ -
still steep compared to $r^{-1}$, but not nearly as steep as $r^{-4}$.}.

The inner shapes of the remnants are typically prolate-triaxial and highly flattened 
($T \sim 0.8$, $\epsilon \sim 0.6$), consistent with previous results for head-on mergers of
stellar systems that have shed most of their orbital angular momentum to the halo 
\citep{ms80,w83,barnes92,tjkin}.  Farther out the remnants become rounder ($\epsilon \sim 0.8$) 
and closer to maximally triaxial.  See also \citet{d94,kaz04,debHaloShapes,vallhalos,abadi09,romdiaz09}, 
who studied halo shapes in a cosmological context and found that the response of the DM to baryonic 
condensation into galaxies drives the halos to rounder and more oblate shapes.

There is substantial cancellation of the prograde and retrograde rotation of the
$z-$tube orbits ($\beta$ near zero), especially in the inner parts.  On the other hand the
$x-$tube populations are highly streaming, with $\beta$ typically approaching one at large
radii.  Remnant $j$ is a notable exception to this rule, with the rotation of its dominant 
$x-$tube orbits canceling nearly perfectly over the full range of radii plotted.  Upon 
closer examination of this remnant, it turns out that this owes to
the symmetry of the ICs.  Both disks on orbit $j$ are substantially inclined, 
but their spins point in opposite directions.  If the $x-$tube orbits in the remnant
are separated by their disk of origin, nearly all of those from disk 1
rotate in one sense about the long axis, while those from disk 2 rotate in the opposite
sense, giving the net cancellation seen in Figure~\ref{fig:rems0}. 

The dissipationless remnants do not show strong orientation twists, and their long axis (which
is typically the axis of symmetry, since the remnants are prolate) lies in the merger 
plane ($\theta_{symm}=90^{\circ}$), along the axis of the final coalescence of the stellar 
components.  This is consistent with the interpretation of \citet{nov06shapes}: the 
time-varying tidal field along the merger axis couples coherently to the stellar 
orbits, producing a bounce that elongates the remnant along the direction of tidal 
compression \citep{mill78,w83}.  The kinematic misalignments are large since the net
rotation tends to favor the long axis, and noisy since the magnitude of the rotation
is often small.

In the $1R_{e}$ kinematic maps, the dissipationless remnants are generally slowly rotating 
(e.g. remnants $j$ and $p$) or are dominated by minor-axis rotation (e.g. remnants $m$ and $n$).  
Some of the remnants (e.g. $p$ and $l$) definitely show both major and minor axis rotation, though 
small in magnitude ($v / \sigma \sim 0.15$).

Figure~\ref{fig:rems0} is a baseline for understanding how a gas component affects the remnants - 
it may be compared with Figures~\ref{fig:rems5} -~\ref{fig:rems40} in the Appendix to see how 
the orbital structure evolves for each merger orbit as $f_{gas}$ is increased from 0 to 40\%.

\begin{figure}[th]
\epsscale{1.17}
\plotone{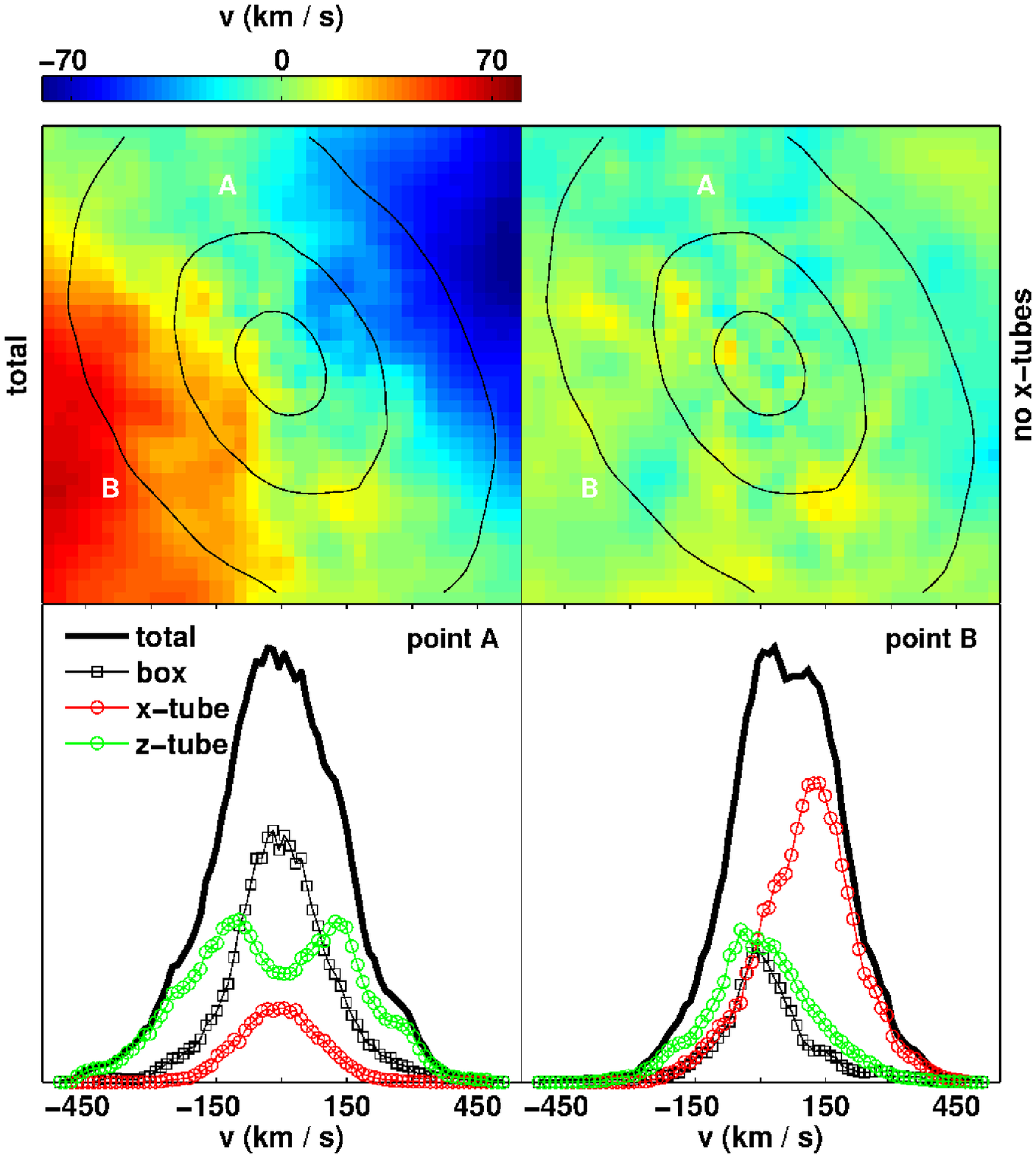}
\caption{{\bf Cancellation of the $z-$tubes and streaming of the $x-$tube orbits.} {\it Upper left:} Velocity map of remnant $m$, 10\% gas, viewed in a near-intermediate axis projection.  {\it Upper right:} Same map with $x-$tube orbits removed.  {\it Lower left:} LOSVD at location A, near the major axis.  {\it Lower right:} LOSVD at location B, near the minor axis.  The velocity scale is the same on both maps.}
\label{fig:cancellation}
\end{figure}

\subsection{Variation with gas fraction}

In Figure~\ref{fig:igfs} we illustrate the effect of dissipation on the remnant structure by varying 
the gas fraction for a fixed merger orbit, $i$, whose structural transformations are representative of 
the overall trends.  Case $i$ is an encounter between a prograde disk in the merger plane 
($\theta_{1}=\phi_{1}=0$), and a highly inclined disk that is also slightly prograde 
($\theta_{2}=71^{\circ}$, $\phi_{2}=30^{\circ}$).  The orbital structure, shape, and orientation
profile of this remnant are shown for seven different gas fractions in Figure~\ref{fig:igfs}.
Unlike in Figure~\ref{fig:rems0}, the velocity maps are all in projection along the $y-$axis,
to clearly show the rotation of both the $x-$tube and $z-$tube orbits and how it varies with $f_{gas}$.

As $f_{gas}$ increases from 0 to 40\%, the population of box orbits within $\sim 1.5R_{e}$
declines, and the $z-$tube population increases.  The most rapid change in the orbital populations
occurs between 0 and 20\% gas.  The outer orbital structure ($r \gtrsim 2R_{e}$) is relatively 
unaffected by the gas, and the $x-$tube population does not display an obvious trend.

The variation in the intrinsic shape closely follows the change in the orbital populations.
As $f_{gas}$ increases, the inner parts become progressively more oblate; the 30-40\% gas 
remnants are nearly oblate axisymmetric within $1R_{e}$ ($T \lesssim 0.1$).  The remnants 
also become rounder with higher $f_{gas}$: $\epsilon$ increases from $\sim$0.6 for the
dissipationless remnant to $\sim$0.8 at the highest gas fractions.

\begin{figure*}[tb]
\epsscale{1.17}
\plotone{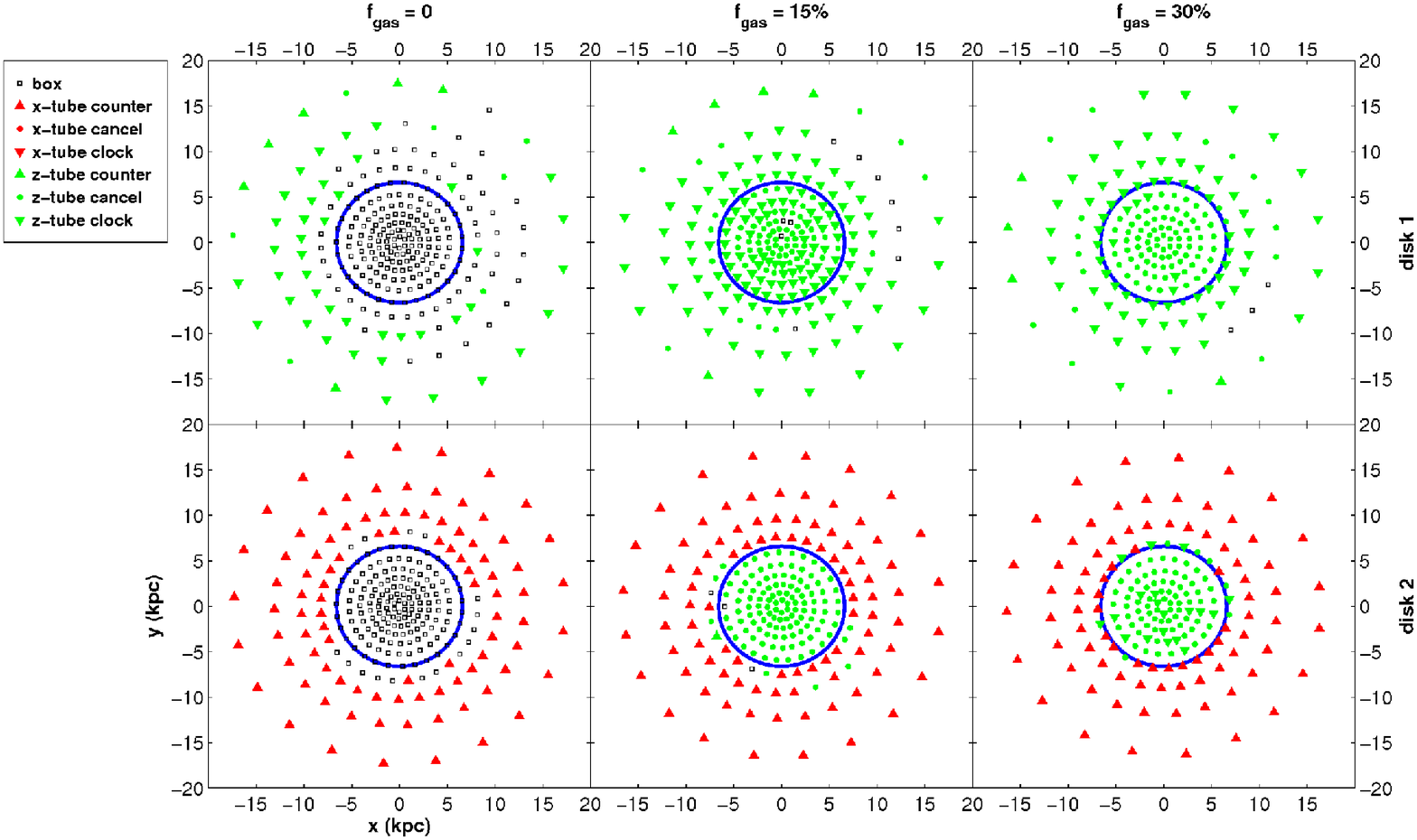}
\caption{{\bf Spatial distribution in the original disks of stars that end up on box, $x-$tube, and $z-$tube orbits, for merger orbit $i$.} Disk 1 (top row) is on a direct, prograde orbit ($\theta_{1} = \phi_{1} = 0$), while disk 2 (bottom row) is inclined and prograde ($\theta_{2}=71^{\circ}$, $\phi_{2}=30^{\circ}$).  Each disk was partitioned into cylindrical bins containing 300 stars each, and we identified the dominant orbital class of the stars from each bin in the final remnant.  If tubes were dominant, then we also computed $\beta$ for the bin.  Bins with $|\beta| < 0.3$ are labeled with circles, while those with $|\beta| > 0.3$ are labeled with triangles pointing up or down according to the sense of rotation (clockwise or counterclockwise, where the overall sign is arbitrary).  The key for different orbital classes is given in the legend.  The three columns show the remnants at 0, 15, and 30\% gas.  Only old stars are included (the gas particles in the initial disks are excluded).  The blue circles mark the half-mass radii of the original disks.}
\vspace{0.1in}
\label{fig:disksc}
\end{figure*}

Though the mass in $z-$tube orbits within $\sim 2R_{e}$ rises substantially as $f_{gas}$ increases 
from 0 to 15\%, the remnants show little change in their major-axis rotation over this range,
since the mass in prograde and retrograde orbits is nearly equal ($\beta \sim 0$).  The 15\%
gas remnant, though dominated by $z-$tubes within 1$R_{e}$, still appears as a minor-axis 
rotator in the projected velocity map.  

In the 20\% gas remnant, the innermost $z-$tube orbits
begin to show substantial streaming, in a sense opposite the net rotation farther out.
This inner streaming shows up as a prominent KDC in the velocity map.  Note that the minor-axis
rotation present at lower gas fractions is still there in this map, but it is dwarfed on the
velocity scale by the rapid rotation within the KDC.  In the 30-40\% gas remnants, the
$z-$tube orbits show significant streaming ($\beta \sim 0.3$) over the full range of radii 
plotted, and the edge-on kinematic maps display rapid, disk-like major-axis rotation.  This
rapid rotation about the intrinsic short axis gives kinematic misalignments near zero 
in the inner parts.

Although the rotation bias of the $z-$tube orbits in the 30\% gas remnant appears fairly uniform, 
the cause of the streaming in the inner and outer parts is quite different.  In the outer
parts the net angular momentum is retained from the ICs, and is therefore
present at all gas fractions.  The inner streaming is a direct consequence of dissipation -
the gas that retains its angular momentum during the merger re-forms a cold disk as it 
dissipates energy, which forms new stars primarily after the violent relaxation of the 
merger is complete.  Thus the inner part of the gas-rich remnants is a superposition of
a hot, slowly-rotating component consisting of the stars present before the merger 
(hereafter referred to as ``old stars''), and a maximally rotating disk composed of
``new stars'' formed through dissipation.  This is why $\beta$ is {\it only} around 0.3 
despite the high maximum velocities in the maps.  We will elaborate on this point in 
section 3.4.

The inner parts of the gas-rich remnants are often rapidly tumbling, which produces large 
intrinsic orientation twists between $\sim$1 and 3$R_{e}$, as in the bottom two rows of
in Figure~\ref{fig:igfs}.  Note that the orientations in the inner parts of these remnants
are an arbitrary snapshot in time; had the simulation been frozen earlier or later $\theta_{symm}$ 
and $\phi_{symm}$ could have quite different values. 

\begin{figure*}[tb]
\epsscale{1.17}
\plotone{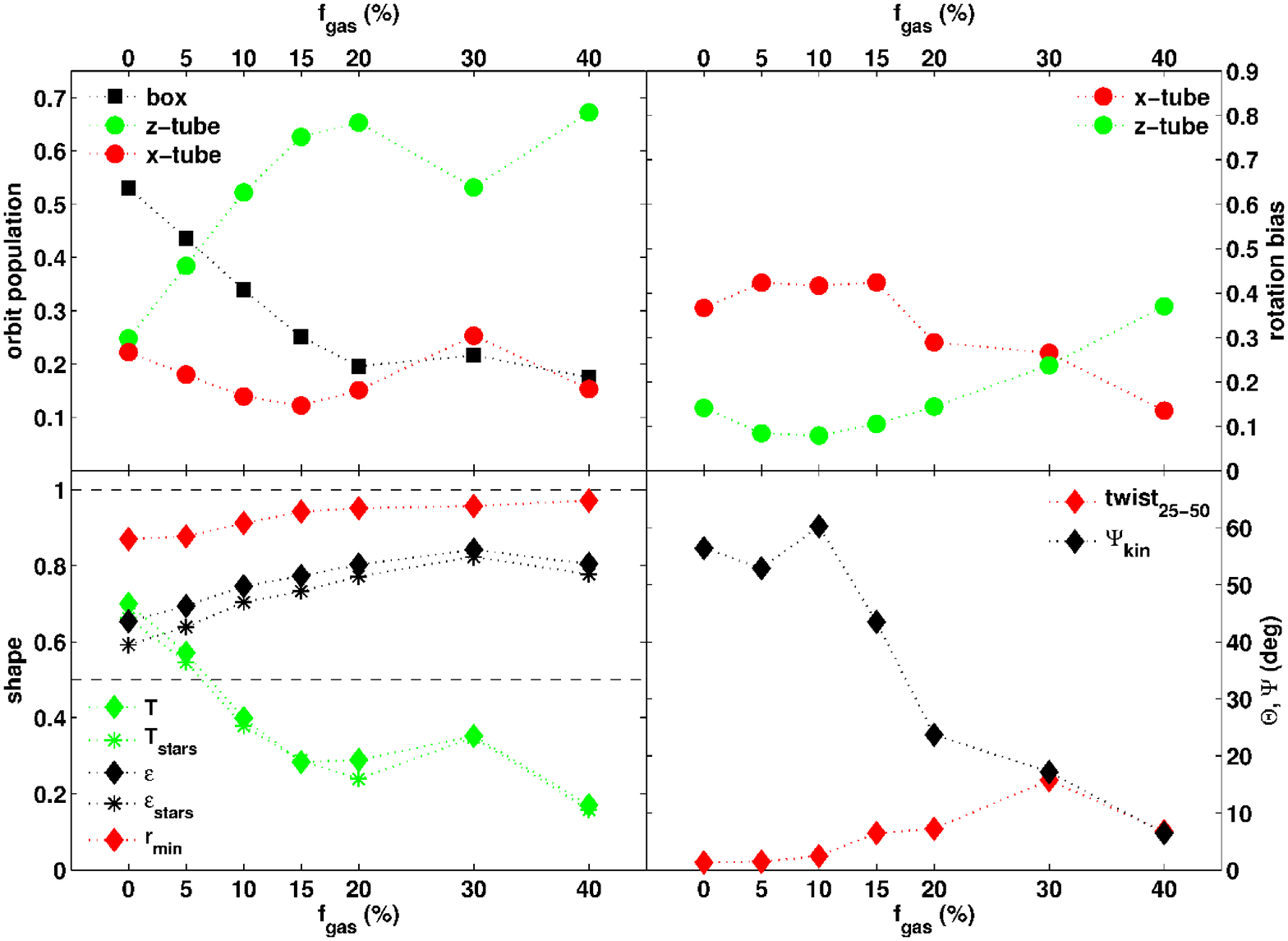}
\caption{{\bf Inner structure ($\lesssim 1R_{e}$), averaged over eight merger orbits.} {\it Upper left:} Fraction of the stellar mass in each orbital class vs. $f_{gas}$, for the 25-45th percentile in binding energy ($\sim 0.5-1.3R_{e}$).  {\it Upper right:} Rotation bias ($\beta$) of the $x-$tube and $z-$tube orbits, as defined in equation 1.  {\it Lower left:} Intrinsic shape of the 40\% most bound particles.  The filled diamonds are for all matter (including DM particles), while the asterisks are for stars only.  {\it Lower right:} Intrinsic orientation twist between the 25th and 50th percentile in binding energy, and kinematic misalignment ($\Psi_{kin}$) for the 40\% most bound particles.  $\Psi_{kin}$ is defined as in Figure~\ref{fig:orbexample}.  The orientation twist is computed as $\max (\theta_{ij})$, $i_{25} < (i,j) < i_{50}$, the maximum angle between the symmetry axes computed within any two cumulative energy bins between the 25th and 50th percentiles.}
\label{fig:inner}
\end{figure*}

In the preceding discussion we have demonstrated that the low-$f_{gas}$ remnants show 
little major-axis rotation not because they lack $z-$tube orbits, but because there is a 
large degree of cancellation between the $z-$tubes rotating in opposite directions.
An observer who measured the major- and minor-axis rotation speeds of the typical 10\% gas
remnant might naively think that the system was prolate in shape and dominated by $x-$tube orbits,
when in fact the predominance of the minor-axis rotation is caused by a sub-dominant population of
$x-$tubes that are highly streaming.  

We highlight this point in Figure~\ref{fig:cancellation} showing remnant $m$ at 10\% gas, which
has comparable masses in $x-$tubes and $z-$tubes within $1R_{e}$, but somewhat more $z-$tubes
(see Figure~\ref{fig:rems10}).  The upper left panel is a velocity map of the remnant in a nearly
edge-on projection (maximizing both the major- and minor-axis rotation), and the second panel
shows the same map with the $x-$tube orbits removed.  Without the $x-$tubes, the remnant shows
almost no rotation at all.  The LOSVDs at the locations marked by the white letters are plotted in
the lower two panels.  The lower left panel clearly shows the canceling streams of $z-$tube orbits
along the major axis, while the lower right panel depicts the highly streaming $x-$tubes along the minor
axis, peaked at nearly 200 km/s.  In the lower left panel, the large population of box orbits fills in
the gap in the bimodal $z-$tube distribution, yielding a total LOSVD that is nearly Gaussian in
shape.

Figure~\ref{fig:disksc} provides insight into the physical explanation for this orbital structure by
mapping out the locations, in the original disks, of the stars that end up in each of the three orbital
classes in remnant $i$, with $f_{gas}=$0, 15, and 30\%.  Disk 1 is the direct disk, and disk 2
is the inclined one.  Only old stars are included in this plot, so the dissipative disk component in
the 30\% gas remnant is left out.  Each disk was partitioned into cylindrical bins containing 300 stars
apiece, and the bins were labeled according to which orbital class formed the plurality of stars among
those 300 in the final remnant - black for boxes, green for $z-$tubes, and red for $x-$tubes.  The
bins favoring tube orbits were further labeled based on whether they tended to stream in one direction
in the final remnant, or to have canceling rotation.  Bins with $\beta > 0.3$ are indicated with
upward-pointing arrows and labeled as ``counterclockwise,'' while those with $\beta < -0.3$ are
indicated with downward-pointing arrows and labeled as ``clockwise.''  Bins with $|\beta| < 0.3$
are labeled as ``canceling'' and indicated with filled circles.  The blue circles mark
the half-mass radii of the original disks.

\begin{figure*}[tb]
\epsscale{1.17}
\plotone{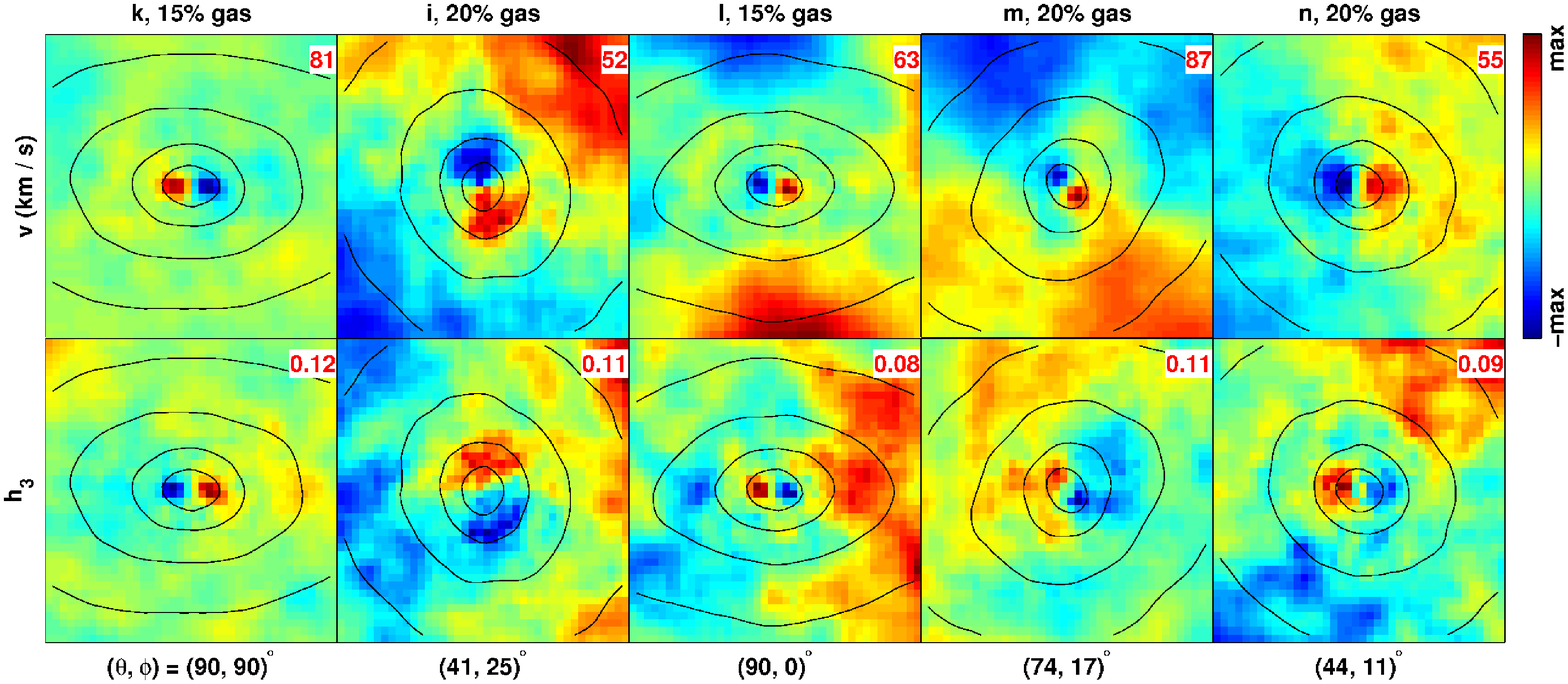}
\caption{{\bf A variety of disk-like KDCs at 15-20\% gas.} {\it Top row:} $1R_{e}$ velocity maps.  {\it Bottom row:} Corresponding $h_{3}$ maps.  The color scaling and labels are as in previous figures.  Each column is labeled at the top with the merger orbit and gas fraction, and at the bottom with the LOS.}
\label{fig:kdcs}
\end{figure*}
\suppressfloats

In the dissipationless simulation, most of the stars within the half-mass radius become
box orbits in the final remnant.  When a dissipative component is added (second and third columns),
the set of stars that became box orbits in the collisionless case instead form a population of $z-$tube orbits
with largely canceling rotation, especially in disk 2.  It is apparent that this cancellation does {\it not}
arise primarily from the ICs, since it occurs for stars in the same disk. Note that the
spatial boundary of the canceling $z-$tube population changes little between 15 and 30\% gas, suggesting
that the boundary is set by a bar formation criterion that determines which orbits ``would have been''
boxes (e.g. \citealt{ms79,ns83,bh96,bour05a,mac06,hop09b}), rather than by the direct gravity of the dissipative
component.  The box population might be fully transformed into $z-$tubes so long as a critical gas mass
is exceeded.  It should also be noted that the efficiency with which the gas is torqued inward
decreases with increasing $f_{gas}$, so the CMC mass comprises a smaller fraction of the initial gas mass
in higher-$f_{gas}$ remnants (see Figure 7 of \citealt{hop09b}).

Stars outside the half-mass radius in disk 1 generally become $z-$tube orbits in the remnant, with a higher
degree of streaming than in the inner parts.  The fate of these stars is relatively insensitive to $f_{gas}$,
though some box orbits from this region in the dissipationless remnant do become $z-$tubes in the simulations
with gas (compare the top right and top middle panels).  The stars originating in the outer part of disk 2
end up in a uniformly streaming $x-$tube population, irrespective of $f_{gas}$.  Their streaming about the
long axis of the remnant can be explained simply in terms of the merger geometry - see e.g. Figure 3 of
\citet{nov06shapes}.  The long axis lies in the merger plane, so stars on nearly circular orbits in the
outskirts of an inclined disk start out with large tangential velocity components about this axis, and are
therefore liable to occupy the $x-$tube ``start space'' (e.g. \citealt{schw79,schw93}).  The uniform rotation
of these orbits in the final remnant suggests that they do not cross orbital boundaries during the merger; their
angular momentum is retained from the initial spin of the inclined disk, and is a direct signature of the
dynamically cold nature of the progenitor galaxy.  Figure~\ref{fig:disksc} provides a good example of how
orbital classification can isolate sets of stars with similar physical histories.

So far we have focused the discussion on individual remnants; Figure~\ref{fig:inner} shows the
dependence of the orbital structure on $f_{gas}$, averaged over all eight merger orbits.  We compute
the structural parameters on $\sim 1R_{e}$ scales since this radius has been the main focus of triaxial dynamical 
modeling efforts to date (e.g. \citealt{schwarzmod,ngc4365,nmagic,vdbthesis}), and defer further discussion of 
the outer orbital structure to section 3.5.

As expected from the preceding discussion, the $z-$tube population increases, the mass in box orbits
decreases, and the shapes become rounder and more oblate with increasing $f_{gas}$.  This transformation
occurs primarily between 0 and 15\% gas.  At higher gas fractions the average shapes, box, and $z-$tube
fractions level off, consistent with the ``critical mass'' interpretation of the
box orbit suppression.  The apparent break in the trend at 30\% gas arises because these
remnants are very round.  In a remnant that is very nearly spherical, the distinction between the 
two classes of tubes is blurred, and the triaxiality parameter and orientation vector are ill-defined
(see e.g. remnant $m$ in Figure~\ref{fig:rems30}).  The 40\% gas remnants are slightly more flattened
than those at 30\% gas, presumably owing to the rotation of the stronger embedded disks.    
The stellar component is slightly more flattened than the mass 
including the DM, particularly at low $f_{gas}$.  There is a mild downward trend in the population of 
$x-$tube orbits with $f_{gas}$ between zero and 15\% as the shape of the potential becomes more oblate,
but the round potentials at higher gas fractions are slightly more amenable to $x-$tubes.

At low $f_{gas}$ the $x-$tube orbits tend to stream far more than the $z-$tube orbits.  This
trend reverses at the highest gas fractions owing to the rapid rotation of the embedded disks, and
the direct modification of the gravitational potential by the large CMC, which may blur the lines
between $x-$tube and $z-$tube orbits as the potential becomes very round.  
The streaming of the $x-$tube orbits might also fall off because of the figure rotation; the
long axis no longer necessarily lies in the merger plane, so more stars may be randomly scattered onto
tangential orbits about the long axis.  The exact reason for this trend in the $x-$tube rotation
needs further investigation.  The cancellation of the $z-$tube rotation is most complete around 
10\% gas, perhaps because the stronger CMC more efficiently disrupts box orbits than at lower 
$f_{gas}$, while at higher $f_{gas}$ the dissipation begins to play a direct role. 

In the high-$f_{gas}$ remnants, the fraction of box orbits often turns upward in the very center. 
This effect is partially hidden in e.g. Figure~\ref{fig:rems40} since we started the profiles
at the 15th percentile in binding energy. It can be naturally explained since only the
innermost component (the CMC itself) contains no smaller CMC, and so can support stable box orbits. 
The violent relaxation during the final merger of the two cores provides a natural setting for the 
onset of triaxiality. However the box population computed within the central starburst is expected
to be highly sensitive to numerical effects, including artificial flattening of the core density 
profile owing to the gravitational softening, and imprecise centering in the orbital classification, 
so we chose to omit the very innermost scales from the structural profiles.

The remnants do not typically show large orientation twists within the 50th percentile in 
binding energy, or $\sim 1.5R_{e}$ (excluding the KDCs, since the shape may not be well-resolved 
on such small scales).  The low $f_{gas}$ remnants typically have large kinematic misalignments,
while the angular momentum of the 40\% gas remnants is well aligned with the short axis of the
potential.

Profiles of the orbital structure, shape, and orientation of all 48 dissipative remnants (5-40\% gas)
are shown in Figures~\ref{fig:rems5} to~\ref{fig:rems40} of the Appendix.  We have outlined 
the main trends with $f_{gas}$ in this section, but there is substantial variation in the structural 
transformations from one merger orbit to the next.  Some of the remnants ($l$ and $p$) are already 
transformed into oblate, $z-$tube dominated systems at 5\% gas, while 
others (e.g. $i$ and $k$) are still box-dominated and substantially prolate at this gas fraction.  
Remnant $o$ is oblate at 0\% gas and becomes prolate at 5\% gas.  A small gas component transforms 
remnant $j$ from a nearly pure prolate spheroid dominated by $x-$tube orbits, to a triaxial system 
with a large $z-$tube population.  As the CMC alters the shape of the potential, both directly and
indirectly through its interactions with the stars, the phase space boundaries between the orbital
classes move. This shifting of the orbital boundaries can induce further changes in the potential,
leading to a cascade effect.  The causes of the more complex orbital transformations in some of the 
individual remnants are an interesting topic for future study.

\subsection{Kinematically distinct cores and embedded disks}

The 15-40\% gas remnants often contain a cold $z-$tube population that appears as a KDC or embedded disk in 
$\sim 1R_{e}$ velocity maps, similar to the sub-components observed in some of the SAURON galaxies 
\citep{sauron12}.

\begin{figure}[t!]
\epsscale{1.17}
\plotone{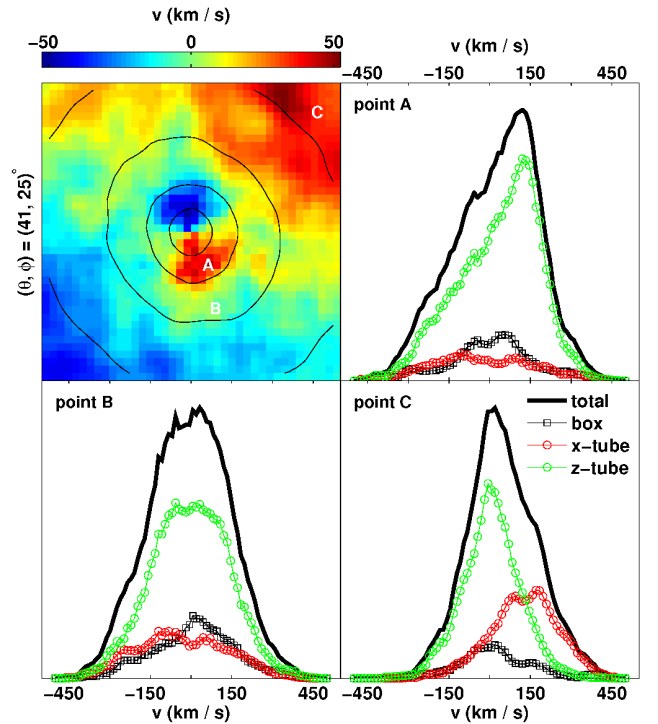}
\caption{{\bf Orbital structure of a typical KDC remnant.} {\it Upper left:} Velocity map of remnant $i$, 20\% gas, viewed in an oblique projection.  {\it Upper right:} LOSVD at location A, near the major axis {\it within} the KDC.  {\it Lower left:} LOSVD at location B, near the major axis {\it outside} the KDC.  {\it Lower right:} LOSVD at location C, near the minor axis.  Color scaling and labels are as in previous figures.}
\label{fig:kdcspec}
\end{figure}

This component typically appears as a KDC in the 15-20\% gas remnants, with a large kinematic twist relative to the 
stars around $1R_{e}$.  Figure~\ref{fig:kdcs} depicts five selected KDC systems, from a
variety of different remnants and viewing angles.  $1R_{e}$ velocity maps are presented in the top row,
while the bottom row shows the corresponding maps of $h_{3}$ to establish the disk-like character of 
the KDCs.  In some cases there are also large photometric twists on the KDC scale, as in column 4 of 
Figure~\ref{fig:kdcs}.  The KDCs are visible from nearly all viewing angles in the $\sim 50$\% of the 
remnants that have them (see Figures~\ref{fig:rems15} -~\ref{fig:rems20}).    

\begin{figure}[b]
\epsscale{1.17}
\plotone{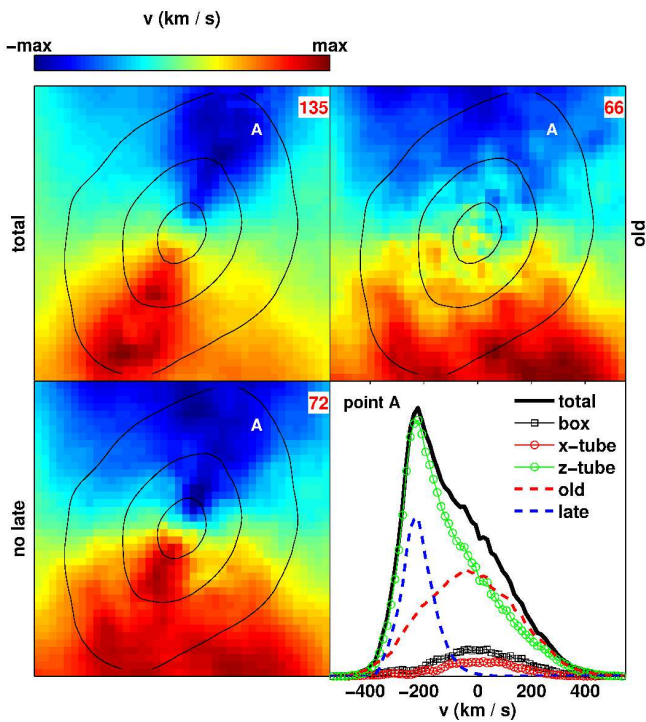}
\caption{{\bf Embedded disk formed by dissipation.}  {\it Upper left:} Velocity map of remnant $i$, 40\% ga\
s, viewed at $(\theta, \phi)=(72, 77)^{\circ}$.  {\it Upper right:} Same map with the new stars, formed fro\
m gas during the merger, removed.  {\it Lower left:} Same map with only the 15\% of the new stars that form\
ed {\it last} removed.  {\it Lower right:} LOSVD at location A, broken down by orbital class and stellar ag\
e.  Color scaling and labels are as in previous figures.}
\label{fig:disk}
\end{figure}

The outer parts of the KDC remnants show a variety of different types of kinematics, depending of the remnant
and projection - some show no rotation (e.g. column 1 of Figure~\ref{fig:kdcs}), some minor-axis or oblique
rotation (columns 2 and 3), and some a significant amount of major-axis rotation either aligned with 
or counter to the KDC (but always less disk-like than the KDC; e.g. columns 2 and 4).
$h_{3}$ and $v$ are always tightly anticorrelated within the KDC, but may have a positive (columns 2 and 5) 
or weaker negative (column 4) correlation outside the core.  Some of the remnants show multiple 
kinematically distinct parts, e.g. remnant $i$ at 20\% gas displays minor-axis rotation with $h_{3}$ and $v$
correlated, and major-axis rotation counter to the KDC, with $h_{3}$ and $v$ anticorrelated, at the outer edge
of the map.

\begin{figure*}[tb]
\epsscale{1.17}
\plotone{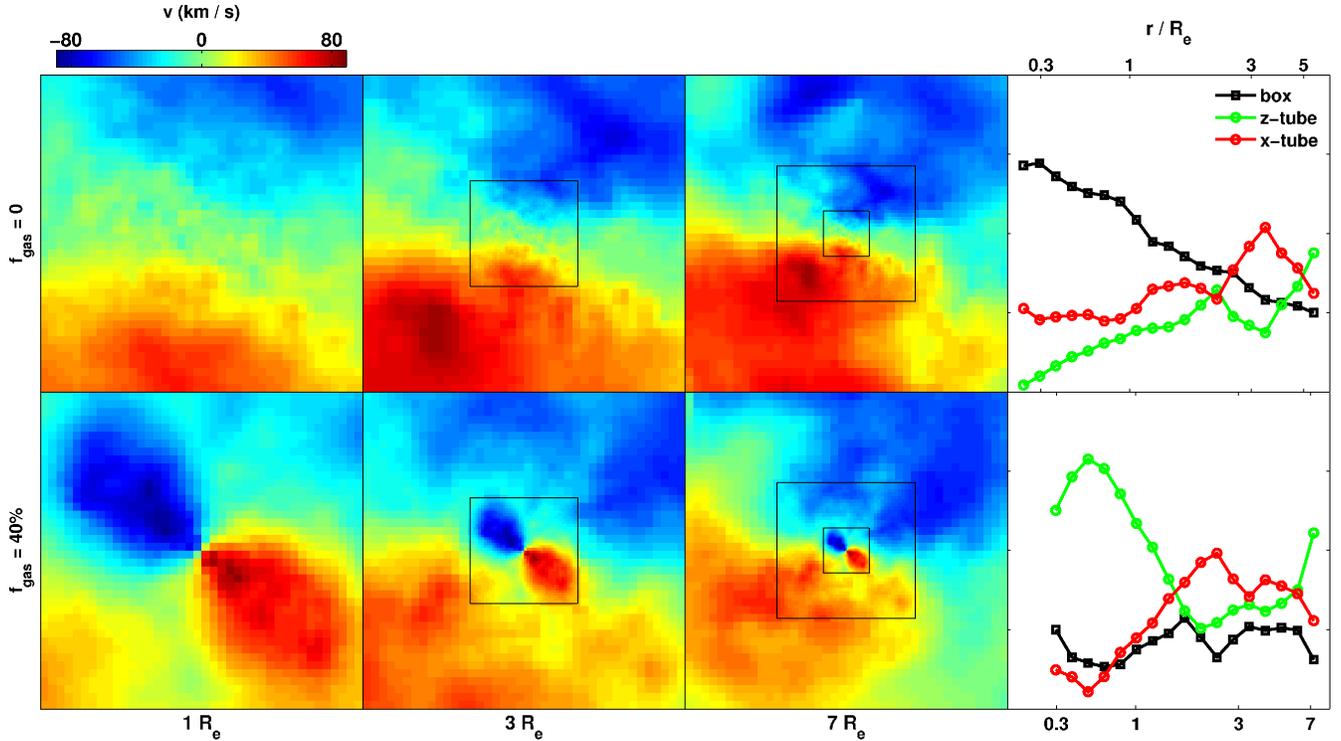}
\caption{{\bf Kinematic transitions in the outer parts.} The top row shows dissipationless remnant $m$, vie\
wed at $(\theta, \phi)=(68, 35)^{\circ}$, while the bottom row shows the same remnant at 40\% gas, along th\
e same LOS.  {\it Left column:} $1R_{e}$ velocity maps.  {\it Left center:} $3R_{e}$ velocity maps, with th\
e $1R_{e}$ maps superimposed inside the boxes.  {\it Right center:} $7R_{e}$ velocity maps, with the 1 and \
$3R_{e}$ maps inside the boxes.  {\it Right:} Radial profiles of the orbital content.  All maps are plotted\
 on the same velocity scale.  Gaussian (rather than GH) fitting was used in this figure because of the nois\
y LOSVDs at large radii.}
\vspace{0.1in}
\label{fig:outerpic}
\end{figure*}

The cores of the KDC remnants are not nearly as distinct in their underlying orbital structure as they are in their 
kinematics, as \citet{ngc4365} observed in the dynamical modeling of NGC4365.  This is because the mass of the 
streaming population producing the KDC is generally small, and the KDCs consist of an oblate $z-$tube population 
on top of a system already oblate and dominated by $z-$tube orbits because of the strong CMC.  Figure~\ref{fig:kdcspec} 
illustrates this point by showing the orbital breakdown of the LOSVDs at three different locations in the KDC remnant 
depicted in the second column of Figure~\ref{fig:kdcs}: one within the KDC (point A), one along the major axis outside 
the KDC, in a region showing no rotation (point B), and one near $1R_{e}$ along the minor axis, in the part dominated 
by minor-axis rotation (point C).  The orbital population at all three locations is (perhaps surprisingly) comprised 
of roughly the same mixture of $z-$tube, $x-$tube, and box orbits, with $z-$tube orbits always in the majority.  The 
very different appearance of the three regions in the kinematic map owes to large differences in the degree of orbital 
streaming - at point A the $z-$tubes are highly streaming; the $x-$tubes are highly streaming at point C; and at point 
B both classes of tubes are distributed symmetrically about $v=0$. 

\begin{figure*}[tb]
\epsscale{1.17}
\plotone{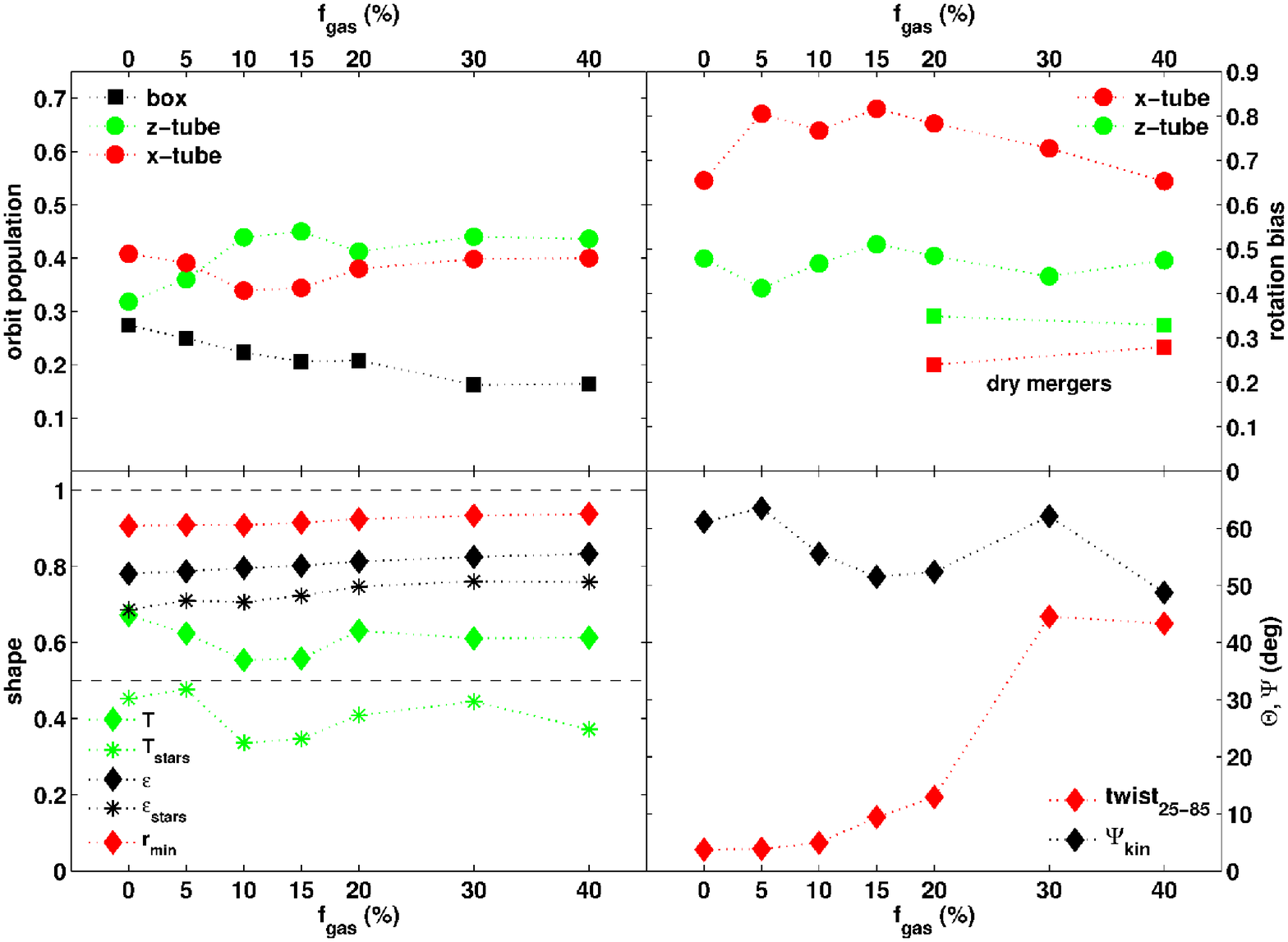}
\caption{{\bf Outer structure ($\sim 3-7R_{e}$), averaged over eight merger orbits.} Same as Figure~\ref{fig:inner}, except that the orbital structure is averaged over the 75-95th percentiles in binding energy, and the shapes and kinematic misalignment are evaluated within the 80th percentile.  The intrinsic orientation twist is computed as $\max (\theta_{ij})$, with $i$ and $j$ running from the 25th to the 85th percentile in binding energy ($\sim 0.5-5R_{e}$).  On the tube rotation plot (upper right panel), we have added data points for re-mergers of the 20 and 40\% gas remnants (labeled as ``dry'' mergers - see text).}
\label{fig:outer}
\end{figure*}

In the 30-40\% gas remnants, the scale of the disk component is typically large enough to fill most or all of the 
SAURON-scale kinematic maps.  The embedded disk in these remnants is typically quite cold, because the disk stars
form in a slow trickle extending over $\sim$2 Gyrs after the merger \citep{r06disks,thesis} and are therefore subject to 
very little violent relaxation.  Figure~\ref{fig:disk} shows remnant $i$ at 40\% gas, which has one of the 
strongest embedded disks in our sample, at a viewing angle about 30$^{\circ}$ from edge on.  The contribution
from old stars, present as collisionless particles in the initial disks, is shown in the upper right panel.
When only old stars are included, the maximum velocity on the map drops from 135 to 66 km/s, and the rotation is no longer
well-aligned with the major axis.  The old stars do still display more major-axis rotation than most remnants at low
$f_{gas}$, since they also contract and spin up in response to the rapid condensation of the gas (e.g.
\citealt{egg62,blum86,gned04,vill09}).  

The lower right panel shows what is left when only the 15\% of the new stars that formed {\it latest} ($\sim$5\% of 
the total stellar mass) are removed.  The maximum velocity still plummets to 72 km/s, and what remains appears 
similar to a KDC system, since the most recently-formed stars dominate the outer portions of the embedded disk.  
The contributions of old and recently-formed stars to a typical LOSVD near $1R_{e}$ in the embedded disk is shown 
in the lower right panel. The stars that formed late have a much smaller velocity dispersion, and when superimposed 
on the hot old population they yield a velocity distribution strongly skewed opposite the mean velocity (see also
\citealt{h3h4}).

\subsection{Outer structure}

We have shown that the outer parts of the remnants (outside $\sim 1.5R_{e}$) are less relaxed 
than the inner parts, and are largely unaffected by the gas content of the disks 
(see Figures~\ref{fig:igfs} and~\ref{fig:disksc}).  While kinematic maps of gas-rich 
remnants are dominated by dissipational rotation on 1$R_{e}$ scales, their kinematics
farther out reflects the ``dry'' part of the merger remnant that is left essentially
untouched by the gas, resulting in large kinematic twists between around 1 and 3$R_{e}$.

This point is illustrated clearly in Figure~\ref{fig:outerpic}, which compares velocity
maps of remnant $m$ at 0 and 40\% gas on three different scales, 1, 3, and 7$R_{e}$,
in the same oblique projection.  Profiles of the orbital populations are reproduced in 
the right-hand panel for comparison.  On 1$R_{e}$ scales the dissipationless and gas-rich 
remnants look entirely different, the former being a minor-axis rotator while the latter
is dominated by rapid disk-like rotation.  However when we zoom out to 3$R_{e}$, the
similarity in their outer structure becomes apparent.  The gas has thoroughly transformed
the inner part of the 40\% gas remnant, while the outer parts are left as they were in the
dissipationless case.  The same observation can be made from the orbital profiles - the
orbital populations outside $\sim 1.5R_{e}$ are similar in the 0 and 40\% gas remnants,
but are radically different within $1R_{e}$, where the dissipationless remnant is comprised
mostly of box orbits and the 40\% gas remnant is dominated by $z-$tubes.  The similarity in 
the outer kinematic maps is even more apparent on 7$R_{e}$ scales.  

Sharp kinematic and orbital transitions such as those in Figure~\ref{fig:outerpic}
will be clearly observable in new surveys such as SMEAGOL, probing the stellar orbital structure
out to $\sim 3R_{e}$ scales \citep{proc09,fost09}, or the PN.S survey, probing the kinematics out 
to $\sim 7R_{e}$ scales using planetary nebulae as tracers \citep{doug02,rom03,cocc09}.  Predictions 
for the statistics of kinematic twists and abrupt changes in the rotation parameter, $\lambda_{R}$ 
\citep{sauron9}, in such surveys will be further discussed and quantified in \citet{hoff09b}. 

The intrinsic outer structure of the remnants is summarized in Figure~\ref{fig:outer}, which is the same
as Figure~\ref{fig:inner} but for the structure between $\sim 3$ and $7R_{e}$.  The orbital
populations, shapes, and orbital streaming are far less sensitive to $f_{gas}$ than in the inner parts.
The outer stellar population is dominated by tube orbits, as expected given the steep effective density
profile at these radii, with a roughly even mix of $x-$tubes and $z-$tubes.  The exchange between box
and $z-$tube orbits with increasing $f_{gas}$ is still noticeable, though diminished.  From visual 
inspection of some randomly selected orbits, box orbits in the halo tend to be more stochastic and spherical 
in shape \citep{gaspard}, and we will further investigate the nature of the outer box orbits in future 
work.

Both classes of tubes have a higher rotation bias than in the inner parts.  The $x-$tube orbits are 
especially highly streaming, which is a signature of the stars' dynamically cold origin.  To highlight 
this point, we have also plotted the streaming fraction of the $x-$tube and $z-$tube orbits over the 
same range of relative binding energies for a series of re-mergers of the 20 and 40\% gas remnants 
(see \citealt{h3h4}), intended to represent ``dry'' mergers between elliptical galaxies that have already 
exhausted their gas in a previous major merger (e.g. \citealt{vd05,bell06,naab06b,k09}).  The streaming 
of the $z-$tubes is not that much different in the remnants of mergers between dynamically hot ellipticals 
than in the disk mergers, but the $x-$tube orbits show far less ordered rotation in the dry merger remnants.  
In a merger between hot systems, the random motions of the stars provide a source of initial angular momentum 
about the long axis, with no preferred sense of rotation.  The uniform streaming of the outer $x-$tube orbits 
might be a tell-tale sign of a cold (late-type) progenitor in dynamical models of observed systems.

The shapes of the matter distribution between 3 and 7$R_{e}$ are round ($\epsilon \sim 0.8$ on average),
and nearly maximally triaxial.  The stars are substantially more flattened ($\epsilon \sim 0.75$) and
oblate ($T \sim 0.4$) than the matter as a whole (including DM).  Large kinematic misalignments are
common in the outer parts, producing large kinematic twists between $\sim$1 and 3$R_{e}$ in the 
most gas-rich remnants, as shown previously in Figure~\ref{fig:outerpic}.  The 30-40\% gas remnants
also typically have large instrinsic orientation twists between $\sim$1 and 5$R_{e}$, owing to the
rapid figure rotation of the inner component.

\section{Summary and conclusions}

\begin{figure}[t]
\epsscale{1.17}
\plotone{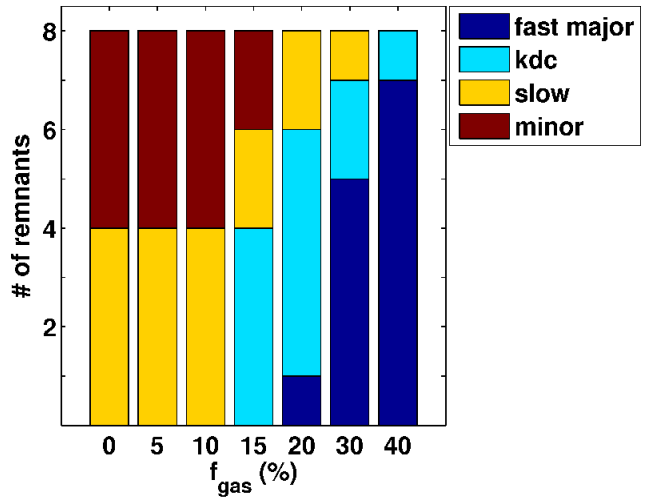}
\caption{{\bf Classification of the remnant kinematics.}  The eight remnants of each gas fraction were categorized as (i) showing rapid major-axis rotation, (ii) slowly rotating with a disk-like KDC, (iii) slowly rotating throughout, or (iv) showing rapid minor-axis rotation, based on visual inspections of their velocity maps (see text).}
\label{fig:kinclass}
\end{figure}

We have shown that a variety of observed kinematic structures can be accounted for 
just by varying the gas fraction in binary 1:1 mergers.  The projected kinematics within 1$R_{e}$
tends to fall into one of four categories:  (i) rapid disk-like rotation about the major axis; 
(ii) a prominent, disk-like KDC with slow or minor-axis rotation farther out; 
(iii) uniform slow rotation, with $v_{max} \lesssim 40$ km/s; and (iv) prominent minor
axis rotation, with little or no rotation about the major axis.  In Figure~\ref{fig:kinclass}
we classify the eight remnants at each $f_{gas}$ into these four categories based on 
visual inspection of their velocity maps in projection along the $y-$axis (intrinsic variations
far outweigh projection effects in the global appearance of the kinematic maps).
The 0-10\% gas remnants are generally slowly rotating or dominated by minor-axis rotation, in agreement with
the previous results of \citet{naab06a,jess07,tjkin}.  The 15-20\% gas remnants often look
similar to the lower-$f_{gas}$ remnants but with disk-like KDCs at their centers, and
bear an interesting resemblance to observed galaxies such as NGC4365 \citep{stat4365,ngc4365,hoff09c},  
NGC5813, and NGC4458 \citep{sauron3,sauron10,vdbthesis}.  

The 30-40\% gas remnants generically show rapid, disk-like rotation with $v/\sigma$ up to $\sim 1$ 
(see also \citealt{rw90,scorz95,r06disks,sauron12}).  None of them display the extreme rotation 
($v/\sigma \sim 2$) of the most rapid rotators in the SAURON sample, although other studies 
\citep{r06disks,robbull08} have shown that 1:1 mergers between 60-80\% gas disks can yield higher
values of $v/\sigma$.  High $v / \sigma$ values are always accompanied by high, anticorrelated
$h_{3}$ values in our remnants, so the subset of SAURON rapid rotators with a shallow $h_{3}-v/\sigma$
relation may require a different formation mechanism.  

The variety of features that can be produced with 1:1 mergers alone suggests that there are 
degeneracies in the formation scenarios leading to a given gross outcome.
For instance cold embedded disks can arise either from gas-rich merging or from
cosmological gas inflow and secular evolution 
\citep{nb01,rober04,coldflows,r06disks,bour07b,robbull08,elme08,gen08SINS,shap08,dekel09,naab09b,kh09,cdb09}. 
Lower progenitor mass ratios ($\sim 3$-4:1 instead of 1:1 mergers) can produce remnants 
resembling many SAURON rapid rotators even in dissipationless simulations 
\citep{naab99,cret01,nb03,bour04,bour05b,burnaab05,jess07,hop09b,jess09,johan09}.  Sequential minor mergers 
can produce slowly rotating ellipticals as well as major mergers with low $f_{gas}$, that are in 
better agreement with observed round, isotropic, and featureless systems such as NGC4486, NGC4552,
and NGC 5846 \citep{wh94,wh95,bour07a,sauron10,bur08,naab09a,gonz09}.  These scenarios must be 
distinguished with more detailed structural and dynamical analysis
(e.g. \citealt{naab06b,jess07,bur08,nmagic2,ngc4365,robbull08,shap08,jeong08,h3h4,th09,decadal,romdiaz09}).

Our orbital analysis suggests a rather simple picture of how the structure
of the remnants arises.  Non-axisymmetric torques in the merger trigger the formation of a
triaxial structure with a large population of box orbits.  When gas is present, the strong
CMC formed through dissipation (e.g. \citealt{mh94a}) converts the majority of the box population
into $z-$tube orbits, which have nearly canceling rotation since box diffusion has no preferred
direction.  Stars in inclined disks begin with a large polar angular momentum component ($j_{\theta}$)
that is retained throughout the merger, so they do not cross orbital boundaries.  This produces
a streaming $x-$tube population in the remnants, that is most pronounced at large radii. 
This picture of the $x-$tube population is supported by the facts that (i) co-planar mergers 
produce remnants almost entirely devoid of $x-$tube orbits; (ii) the phase space available to 
$x-$tubes in the remnant potentials is generally underpopulated \citep{barnes92,thesis}; and
(iii) in some observed ellipticals the velocity ellipsoid is flattened along the polar axis 
\citep{vdbthesis}.

In gas-rich merger remnants, there is also a pronounced streaming $z-$tube population on small
scales owing to the gas that dissipates energy and retains its angular momentum during the
merger \citep{hop09b}.  Outside a fairly well-defined boundary (see e.g. Figures~\ref{fig:disksc} 
and~\ref{fig:outerpic}), the orbital structure is largely unaffected by the gas.  This boundary
corresponds to the radius where box orbits cease to dominate in the dissipationless remnants,
implying that these regions may be left relatively intact because they have never had a large population
of radial orbits to convey information about the galactic center to their location.  Time-dependent
orbital classification and idealized evolving models are needed to verify this interpretation 
(e.g. \citealt{hoff09c,vill09}). 

The characteristic orbital structure arising from this picture, and in particular the sharp kinematic
transitions predicted between $\sim 1$ and $3R_{e}$, should be readily observable with surveys in
progress such as SMEAGOL.  Some pronounced kinematic transitions have been observed around these radii
in survey pilot studies \citep{proc09,cocc09}, but more data will be needed to determine whether these
transitions are of the same nature as those in the merger simulations.  Note that some features, e.g. 
the uniform streaming of the $x-$tube orbits, can only be captured with dynamical modeling - substantial 
minor-axis rotation alone could be produced by either a dominant population of $x-$tubes with a small 
rotation bias, or a smaller, highly streaming population. 

We do not necessarily expect most real galaxies to display the sharp features in their orbital
structure found in this paper, since cosmological galaxy formation histories are far more complex 
than isolated 1:1 mergers between pure disks.  However the results of our analysis illustrate
how orbital analysis can isolate subsets of the stellar population with similar histories
and place intuitive constraints on galaxy formation mechanisms.  In future work we hope to
extend this type of analysis to a broader range of merger parameters and more complex 
cosmological formation scenarios.

%\\[0.25in]

\begin{acknowledgments}
We thank Glenn van de Ven, Remco van den Bosch, Aaron Romanowsky, and Phil Hopkins for 
enlightening discussions, especially on relating our work to observations.  We are also 
grateful to Bart Willems and Jaczek Braden for technical help.  This work was supported in 
part by a Lindheimer Postdoctoral Fellowship at Northwestern University.  Computations were 
performed on the {\it Fugu} computer cluster at Northwestern, funded by NSF MRI grant PHY-0619274 
to Vicky Kalogera, and the {\it Sauron} cluster at the parallel computing center of the 
Institute for Theory and Computation at the Harvard-Smithsonian Center for Astrophysics.
\end{acknowledgments}

\vspace{2in}

\appendix

%\addtocounter{figure}{14}

\begin{figure*}[htb]
\epsscale{1.2}
\plotone{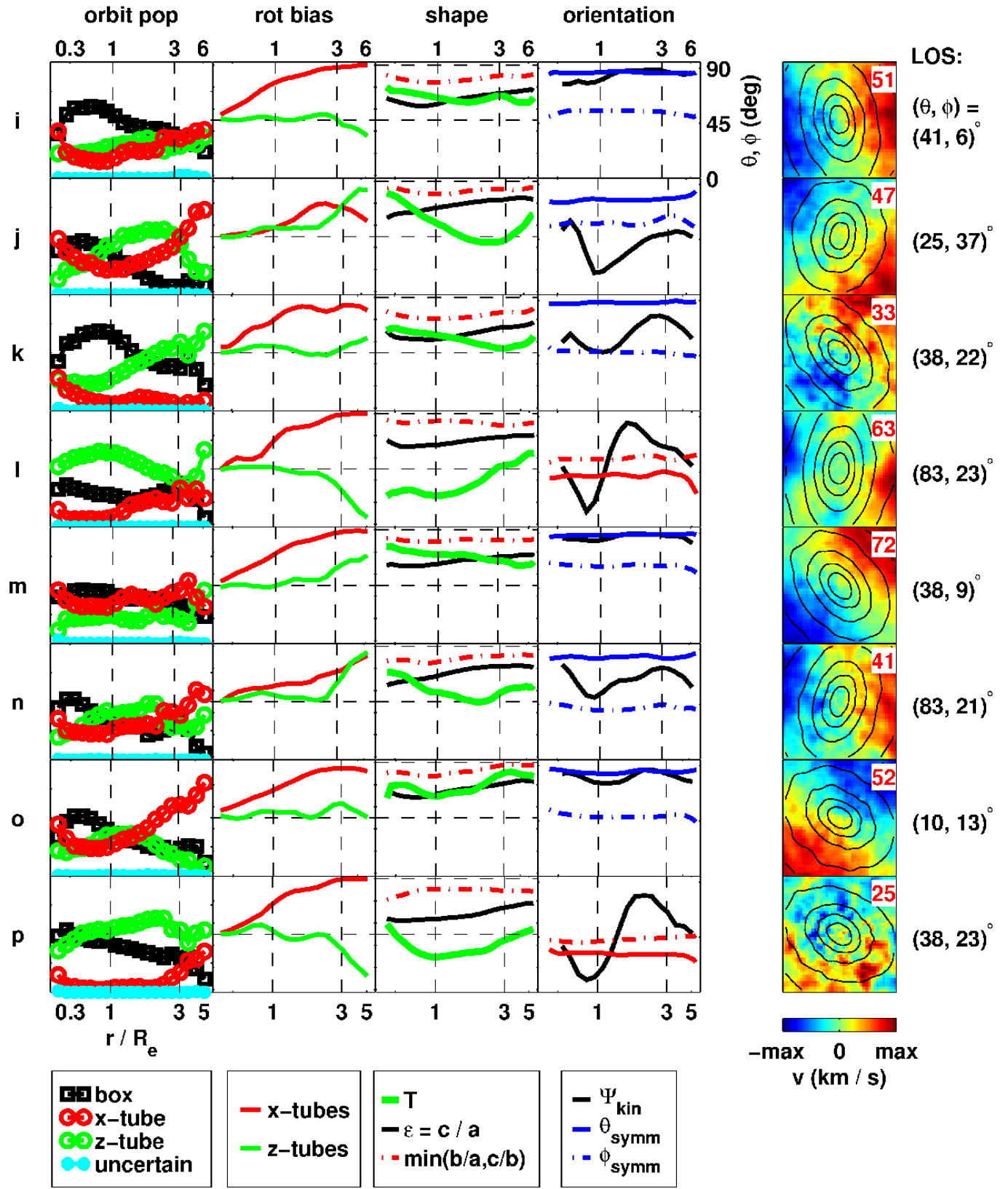}
\caption{{\bf Intrinsic structure of the 5\% gas remnants.}}
\label{fig:rems5}
\end{figure*}

\begin{figure*}[htb]
\epsscale{1.2}
\plotone{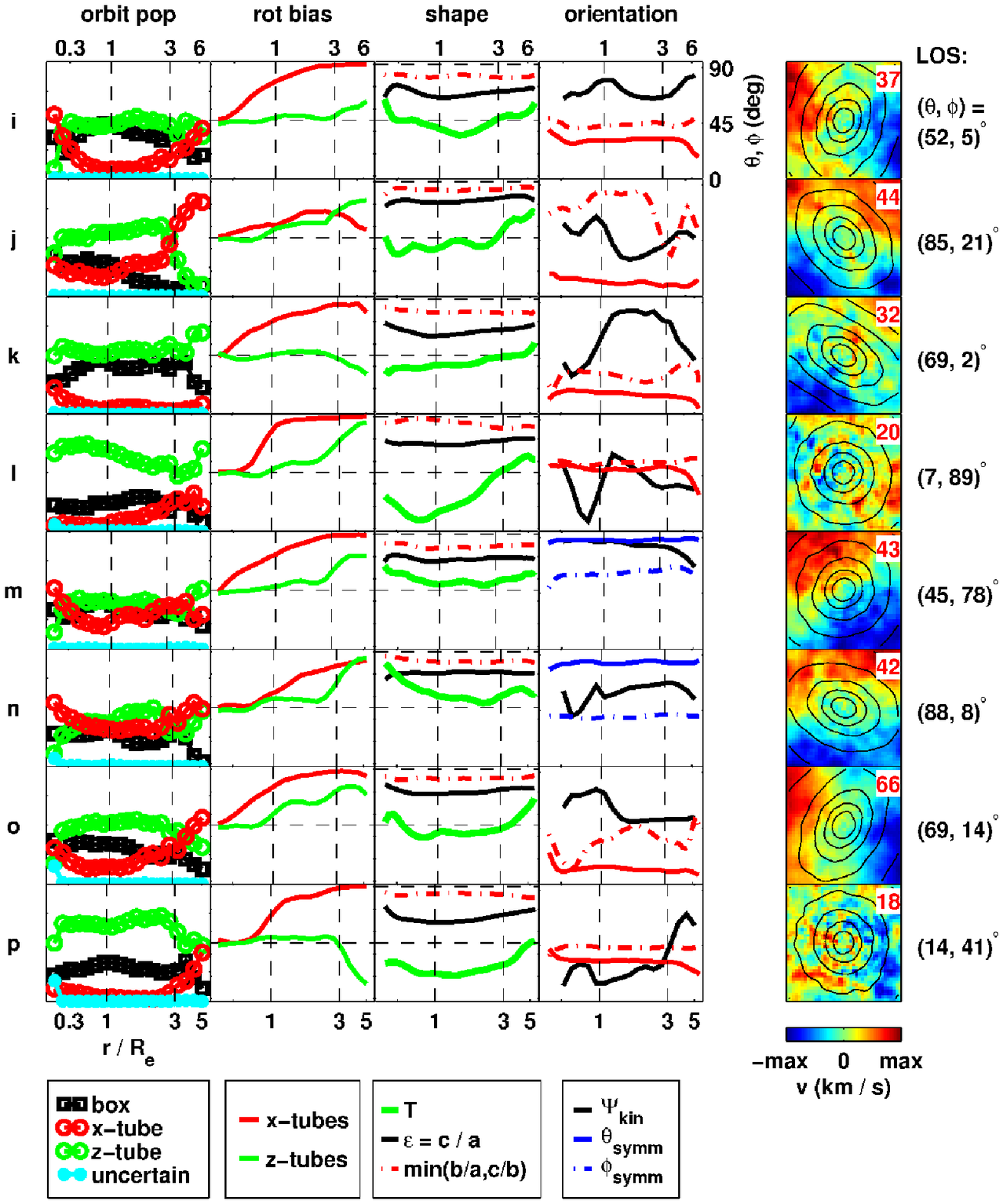}
\caption{{\bf Intrinsic structure of the 10\% gas remnants.}}
\label{fig:rems10}
\end{figure*}

\begin{figure*}[htb]
\epsscale{1.2}
\plotone{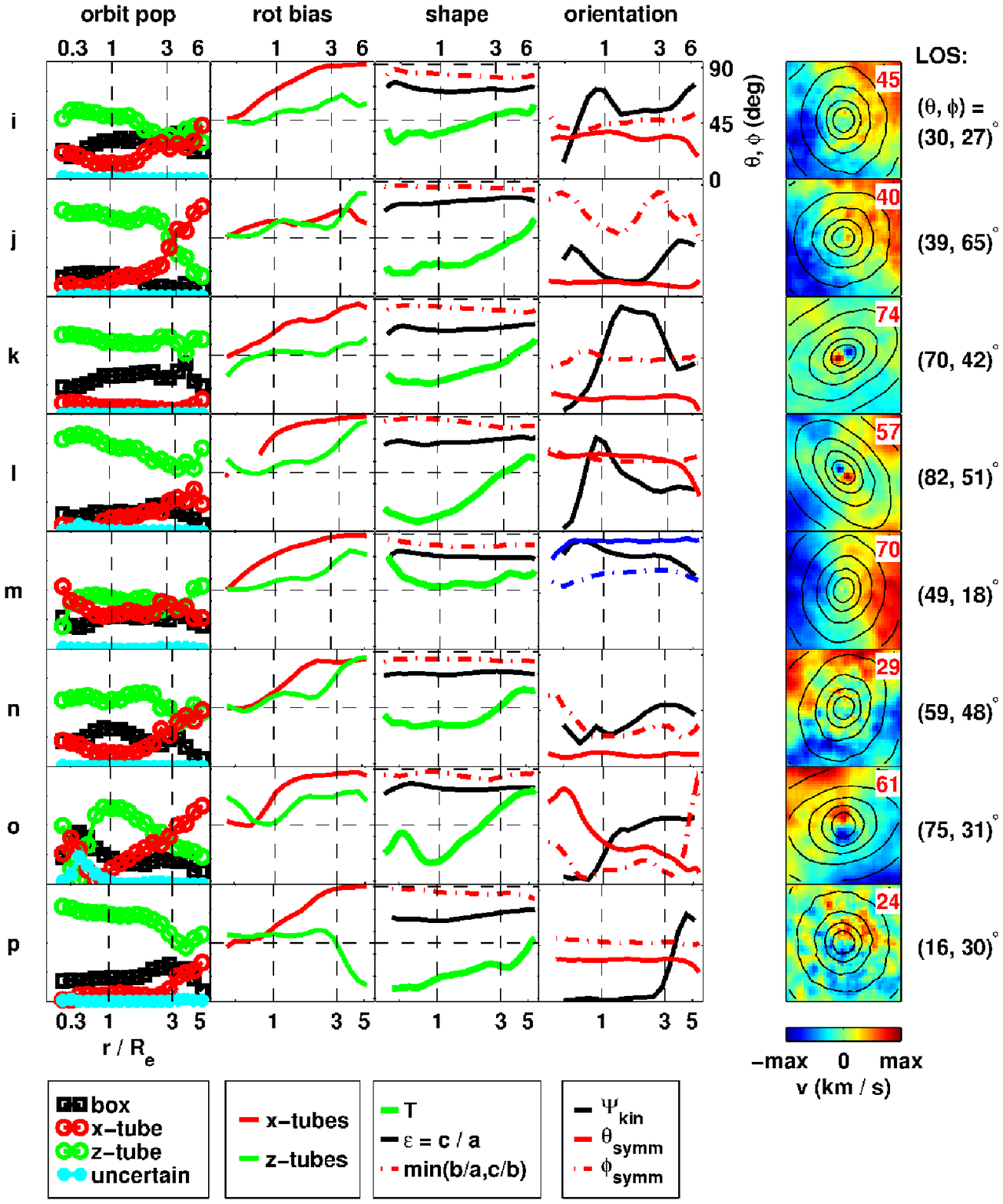}
\caption{{\bf Intrinsic structure of the 15\% gas remnants.}}
\label{fig:rems15}
\end{figure*}

\begin{figure*}[htb]
\epsscale{1.2}
\plotone{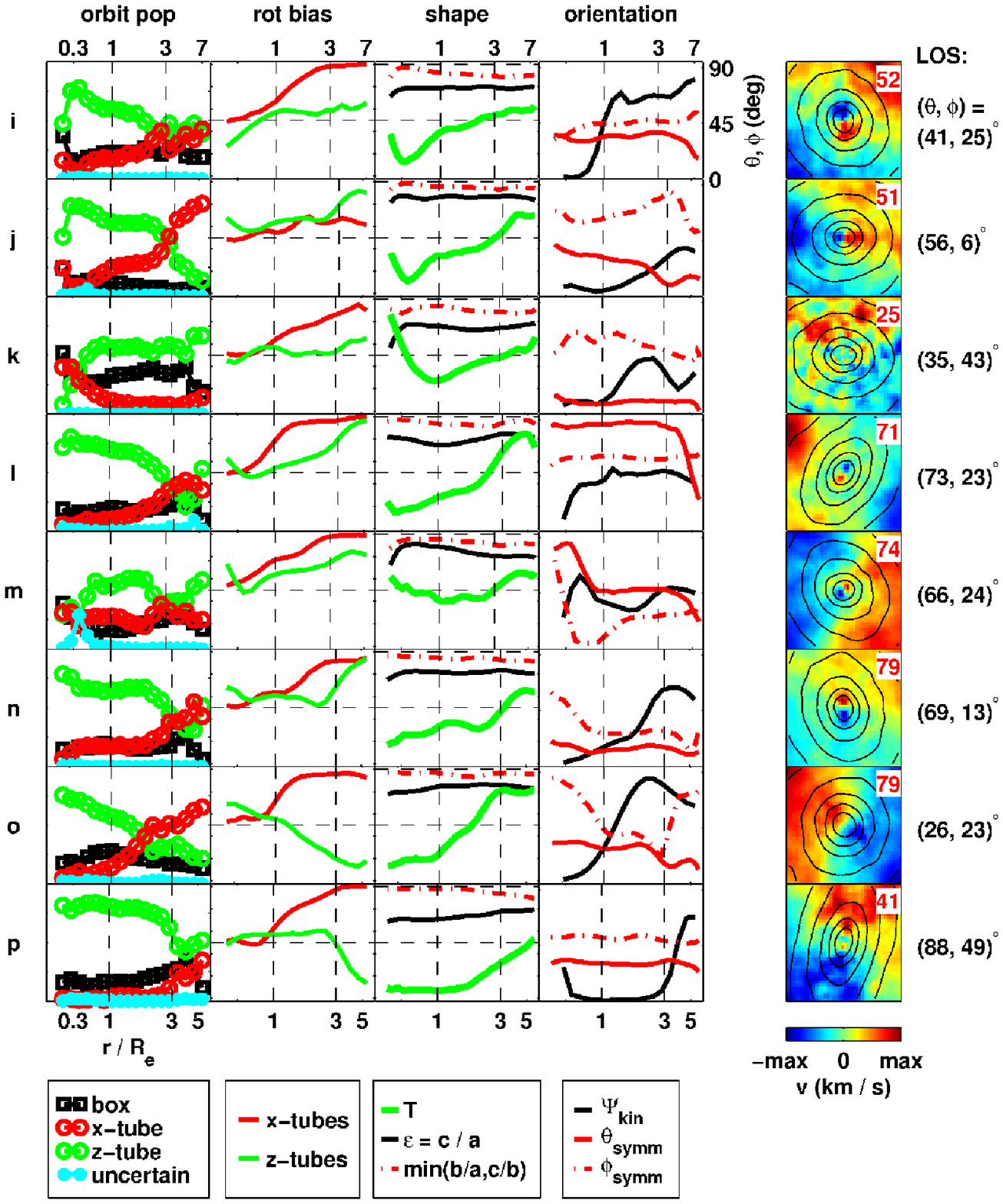}
\caption{{\bf Intrinsic structure of the 20\% gas remnants.}}
\label{fig:rems20}
\end{figure*}

\begin{figure*}[htb]
\epsscale{1.2}
\plotone{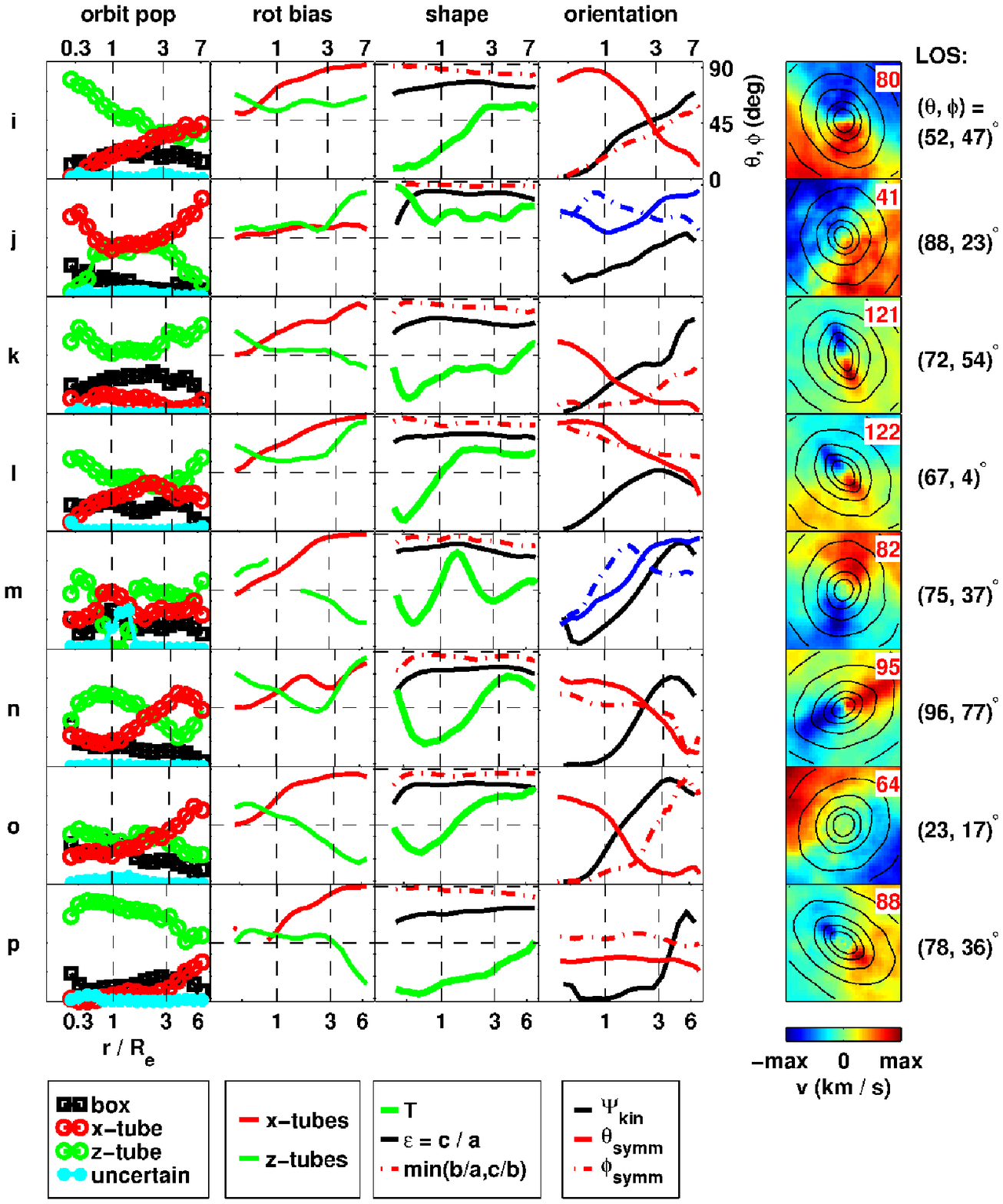}
\caption{{\bf Intrinsic structure of the 30\% gas remnants.}}
\label{fig:rems30}
\end{figure*}

\begin{figure*}[htb]
\epsscale{1.2}
\plotone{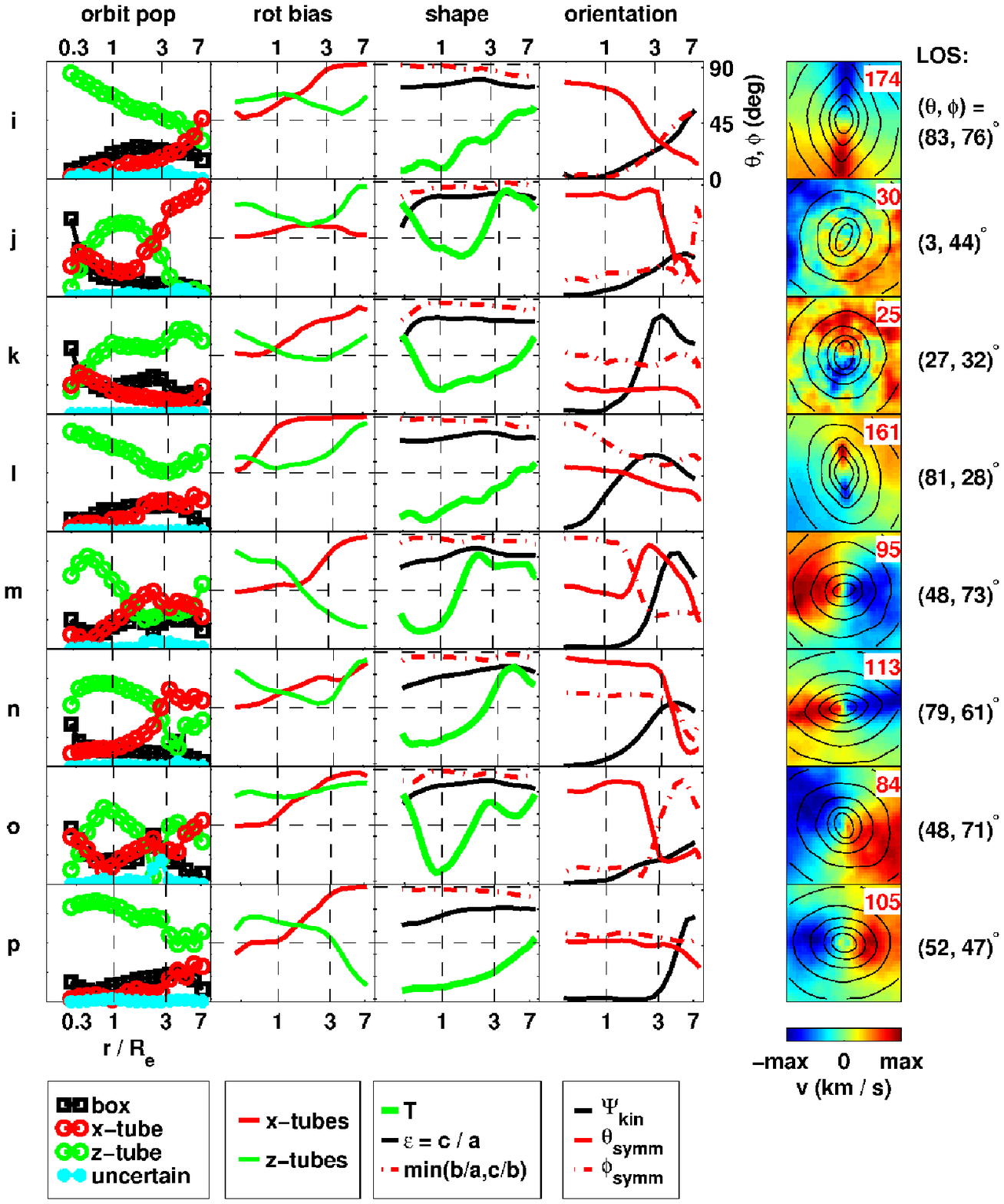}
\caption{{\bf Intrinsic structure of the 40\% gas remnants.}}
\label{fig:rems40}
\end{figure*}

In Figures~\ref{fig:rems5} -~\ref{fig:rems40} we present radial profiles of the orbital populations, intrinsic shapes, and 
orientations of all eight merger remnants with each initial gas fraction (5, 10, 15, 20, 30, and 40\%).  The format of the
figures is identical to that of Figure~\ref{fig:rems0} in the body of the paper (instrinsic structure of the dissipationless
remnants).  The figures may be viewed one-at-a-time to get a feel for the variation with merger orbit at fixed $f_{gas}$,
or compared row-for-row to follow the evolution of the structure with $f_{gas}$ for a fixed merger orbit.

\end{document}